\documentclass[useAMS,usenatbib]{mn2e}

\usepackage{verbatim,graphicx,epsfig,dcolumn,url}
\newcolumntype{.}{D{.}{.}{4}}
\newcolumntype{,}{D{.}{.}{2}}
\newcolumntype{;}{D{.}{.}{1}}
\newcommand{\nodata}{$\cdot\cdot\cdot$}
\newcommand{\lesssim}{{\lower-1.2pt\vbox{\hbox{\rlap{$<$}\lower5pt\vbox{\hbox{$\sim$}}}}}}
\newcommand{\gtrsim}{{\lower-1.2pt\vbox{\hbox{\rlap{$>$}\lower5pt\vbox{\hbox{$\sim$}}}}}}

\def\clap#1{\hbox to0pt{\hss#1\hss}}

\title[Fundamental parameters and infrared excesses from Hipparcos]{Fundamental parameters and infrared excesses of Hipparcos stars}
\author[I. McDonald et al.]{I.~McDonald$^{1}$\thanks{E-mail:
mcdonald@jb.man.ac.uk}, A.~A.~Zijlstra$^{1}$, M.~L.~Boyer$^{2}$\\
$^{1}$Jodrell Bank Centre for Astrophysics, Alan Turing Building, Manchester, M13 9PL, UK\\
$^{2}$STScI, 3700 San Martin Drive, Baltimore, MD 21218, USA}

\begin{document}

\date{Accepted 9999 December 32. Received 9999 December 32; in original form 9999 December 32}

\pagerange{\pageref{firstpage}--\pageref{lastpage}} \pubyear{9999}

\maketitle

\label{firstpage}

\begin{abstract}
We derive the fundamental parameters (temperature and luminosity) of 107\,619 \emph{Hipparcos} stars and place these stars on a true Hertzsprung--Russell diagram. This is achieved by comparing {\sc BT-Settl} model atmospheres to spectral energy distributions (SEDs) created from \emph{Hipparcos}, \emph{Tycho}, SDSS, DENIS, 2MASS, \emph{MSX}, \emph{AKARI}, \emph{IRAS} and \emph{WISE} data. We also identify and quantify from these SEDs any infrared excesses attributable to circumstellar matter. We compare our results to known types of objects, focussing on the giant branch stars. Giant star dust production (as traced by infrared excess) is found to start in earnest around 680 L$_\odot$.
\end{abstract}

\begin{keywords}
stars: fundamental parameters --- solar neighbourhood --- stars: mass-loss --- circumstellar matter ---  infrared: stars --- Hertzsprung--Russell and colour--magnitude diagrams
\end{keywords}


\section{Introduction}
\label{IntroSect}

Spectral energy distributions (SEDs) have long been the primary method of understanding stars. Colour--magnitude diagrams, which can be quickly made from photometric data, enable one to explore various facets of stellar populations, such as stellar mass and evolutioanry state. However, these does not present the information at its most basic physical level: the stellar temperature and luminosity. These represent the fundamental ideals of stellar modelling, and are theoretically free from biases introduced by photometric calibration, interstellar reddening and similar phenomena.

While transformations to these parameters can be achieved through colour--temperature relations and bolometric corrections, these are limited in scope. Most importantly, the wavelength coverage of the observations means that well-defined solutions do not always exist for these relations (e.g. for very red stars, or for observations only covering wavelengths longer than the SED peak). Using the entire wavelength coverage available allows better determination of stellar temperature when a wide temperature range is present among a stellar sample. This also allows finer control of data quality. All-sky surveys are, in particular, prone to contain some poor-quality data due to the large flux range they are required to cover, which leads to the saturation of bright sources, and the volume of data, which limits the ability to match photometric routines to particular situations (e.g.\ in areas of high stellar density or nebular emission). Stellar variability can also cause improper colours to be reported, which can be reduced by using multiple epochs or, equivalently, multi-wavelength data. In this manner, we can provide more-robust estimates of parameters for individual objects, allowing them to be placed on the true, physical Hertzsprung--Russell (H--R) diagram.

Perhaps the greatest benefit, however, is the ability to detect excess flux at a particular wavelength, by providing a reference model flux against which fluxes in individual photometric filters can be compared. This is particularly helpful in the infrared, where colour--magnitude diagrams based on only part of the SED can fail to identify sources exhibiting emission in addition to the stellar photosphere. Predominantly, these sources are either very young stars (pre-main-sequence T Tauri stars or Herbig Ae/Be stars), rapid rotators (classical Be stars), or evolved stars. This latter group is mostly comprised of mass-losing red and asymptotic giant branch (RGB/AGB) stars, on which we focus our discussion.

Previously, only colour--magnitude diagrams have been used to interpret our closest stellar neighbours (e.g.\ \citealt{PLK+95}). We are now able to take the data returned by the \emph{Hipparcos} satellite \citep{Perryman97,vanLeeuwen07} and match it with other all-sky surveys to produce a true H--R diagram of the local Solar neighbourhood.

In doing so, we can identify and characterise stars with weak infrared excesses which may be otherwise missed by conventional colour cuts. While this has been attempted previously (\citealt{IMI+10,Groenewegen12}; we later discuss these papers in context), this work represents the first time such a process has been applied to the entire \emph{Hipparcos} dataset and in the context of the stars' absolute, fundamental parameters.


\section{Fundamental stellar parameters}
\label{SEDSect}

\subsection{Input data catalogue}
\label{DataSect}

\begin{center}
\begin{table*}
\caption{Number of sources used from each catalogue.}
\label{CatsTable}
\begin{tabular}{l@{\qquad\qquad}r@{\qquad\qquad}r@{\qquad\qquad}r@{\qquad}r@{\qquad}r}
    \hline \hline
Catalogue	& Wavelength	& Beam size	& \multicolumn{3}{c}{Sources used} \\
\ 		& ($\mu$m)	& ($^{\prime\prime}$)& Original	& Combined	& Final\\
\ 		& \ 		& \ 		& data		& catalogue	& catalogue\\
    \hline
\emph{Hipparcos}& 0.528	& $\sim$0.5	& 117\,956	& 109\,661	& 107\,616	\\
\emph{Tycho}	& 0.420, 0.532	& $\sim$0.5	& 118\,924	& 109\,624	& 107\,586	\\
SDSS		& 0.354--0.623	& $\gtrsim$0.5	&  32\,253	&  30\,368	&  27\,420	\\
DENIS		& 0.786--2.20	& $\gtrsim$0.5	&  60\,083	&   2\,856	&   2\,762	\\
2MASS		& 1.25--2.20	& $\gtrsim$0.5 	& 104\,324	& 104\,297	& 104\,111	\\
\emph{MSX}	& 4.29--21.3	& $\approx$18.3 &   7\,663	&   7\,336	&   3\,153	\\
\emph{AKARI}	& 10.5, 18.4	& 2.4, 2.3 	&  48\,078	&  48\,013	&  47\,762	\\
\emph{IRAS}	& 12, 25	& 106, 106	&  19\,728	&  16\,001	&  15\,533	\\
\emph{WISE}	& 3.35--22.1	& 5.8--11.8	&  64\,192	&  64\,102	&  63\,883	\\
    \hline
\multicolumn{6}{p{0.7\textwidth}}{Notes: The ground-based optical beam sizes of SDSS, DENIS and 2MASS are limited by seeing, hence presented as approximate lower limits. Beam sizes of scanning satellites are not circular: here, the equivalent-sized circular aperture is given instead. The combined catalogue contains all \emph{Hipparcos} objects for which we were able to obtain any matching infrared data. The final catalogue contains only the objects remaining after the data quality cuts described in the Appendix had been carried out.} \\
    \hline
\end{tabular}
\end{table*}
\end{center}

The new \emph{Hipparcos} ($H_{\rm p}$) / \emph{Tycho} ($B_T$, $V_T$) reduction \citep{vanLeeuwen07} was used as the primary astrometric and photometric catalogue, to which the other catalogues were matched. Additional data was sourced from the following surveys:
\begin{list}{\labelitemi}{\leftmargin=1em \itemsep=0pt}
\item Sloan Digital Sky Survey (SDSS-III) Data Release 8 \citep{AAPA+11}: $ugr$-band optical data were included. These data are heavily affected by saturation for the \emph{Hipparcos} sample. The $iz$-band photometry were left out entirely, and bad data from the $ugr$-bands were identified as described in \S\ref{BadSect}.
\item DENIS Consortium 2005 data release\footnote{VizieR On-line Data Catalog: II/263}: $i_{\rm Gunn}JK_{\rm s}$-band data were included for sources where $i_{\rm Gunn} > 9.5$ mag. $JK_{\rm s}$-band near-IR data were used only when 2MASS data were not available (see below).
\item 2MASS All-Sky Catalog of Point Sources \citep{SCS+06}: $JHK_{\rm s}$-band near-IR data were included. All photometry was initially included, regardless of data quality: bad data was later removed as described in \S\ref{BadSect}.
\item \emph{MSX} Infrared Astrometric Catalog \citep{EP96}, incorporating six bands ($B_1$, $B_2$, $A$, $C$, $D$ and $E$) covering 4--18 $\mu$m. Note that, at $\approx$18$^{\prime\prime}$ resolution, these data exhibit problems from source blending.
\item The \emph{AKARI--Hipparcos} cross-correlated catalogue \citep{IMI+10}, covering 9 and 18 $\mu$m, hereafter \emph{AKARI} [9] and [18].
\item \emph{IRAS} catalogue of Point Sources, Version 2.0 (PSC; \citealt{BHW88}), and Faint Source Catalog, $|b|>10$, Version 2.0 (FSC; \citealt{MCC+90}), both limited to the 12- and 25-$\mu$m bands, hereafter \emph{IRAS} [12] and [25]. At $\approx$1$^{\prime}$ resolution, these data also exhibit problems from source blending.
\item The \emph{WISE} Premiminary Data Release \citep{WEM+10}, incorporating four bands ($W_1$ through $W_4$) covering 3.5--22 $\mu$m.
\end{list}

As the surveys cover a large range of wavelengths and have very differently-sized point-spread functions (PSFs; Table \ref{CatsTable}), one must be conservative in declaring two detections as arising from the same source. DENIS, 2MASS and \emph{AKARI} detections were considered to be matched with \emph{Hipparcos} detections when their co-ordinates matched to within 1$^{\prime\prime}$, SDSS and \emph{WISE} data were matched when within 3$^{\prime\prime}$, and \emph{MSX} and \emph{IRAS} data were matched within 5$^{\prime\prime}$.

Due to the differing depths and similar wavelength coverages of the near- and mid-IR catalogues, substitutions were made to choose only the most discerning data. Specifically, where possible, \emph{MSX} and \emph{IRAS} data were replaced by higher-resolution, greater-sensitivity \emph{WISE} and \emph{AKARI} data; also DENIS photometry was replaced by 2MASS photometry, which is less prone to saturation. These substitutions were performed as follows:
\begin{list}{\labelitemi}{\leftmargin=1em \itemsep=0pt}
\item $W_1$ and $W_2$ replace \emph{MSX} $B_1$ and $B_2$, respectively;
\item \emph{AKARI} [9] replaces \emph{MSX} $A$;
\item $W_3$ and/or \emph{IRAS} [12] replaces \emph{MSX} $C$ and $D$;
\item $W_4$ and/or \emph{AKARI} [18] replaces \emph{MSX} $E$;
\item 2MASS $J$ and $K_{\rm s}$ replaces DENIS $J$ and $K_{\rm s}$;
\item \emph{IRAS} FSC data replaces \emph{IRAS} PSC data.
\end{list}
SDSS bands were dropped under certain conditions, namely:
\begin{list}{\labelitemi}{\leftmargin=1em \itemsep=0pt}
\item $u$ was dropped if $u > B_T + (B_T - V_T) + 2$ mag;
\item $g$ was dropped if $g > B_T + 2$ mag;
\item $r$ was dropped if $r > V_T + 2$ mag;
\end{list}
which correspond to ranges beyond which the SDSS data (which is prone to saturation for the \emph{Hipparcos} stars, most of which are comparatively bright) cannot be matched in simultaneity with the \emph{Tycho} data to any stellar model.

The resulting initial input catalogue contains 109\,661 \emph{Hipparcos} stars with data from $u$-band to 25 $\mu$m. The source statistics for this combined catalogue are listed in Table \ref{CatsTable}.

\subsection{Fitting the SEDs}
\label{FitSect}

The SEDs were fitted using the code initially described in \citet{MvLD+09} and modified in the subsequent papers \citep{BMvL+09,MSZ+10,MvLDB10,WOK+11,MBvL+11,MJZ11,MvLS+11}. This code, hereafter referred to as {\sc Getsed}, is optimised to detect low-contrast infrared excess arising from circumstellar dust. We have made some further revisions to the code as detailed below.

{\sc Getsed} works by minimising the $\chi^2$ statistic between the observed photometric data and a set of synthetic stellar spectra to determine stellar temperature and luminosity. This requires user-defined stellar mass, metallicity and distance, and an interstellar reddening, and given appropriate filter transmission curves. A grid of temperatures between 2400 and 60\,000 K is set up, in steps of 400 K. Blackbodies of these temperatures are fit to the dereddened photometric data and the model with the lowest $\chi^2$ chosen. A finer grid is set up and the blackbody temperature is iterated to the nearest 25 K. This temperature is then used to derive the stellar luminosity and surface gravity to first order.

For this work, we do not know the mass or metallicity of our stars, which are required to select the correct set of synthetic spectra. Nor do we know the interstellar reddening toward the stars. We assume that the metallicity is solar and that the interstellar reddening is zero. An assumption of solar metallicity is reasonable for nearby Galactic stars: the true values scatter around the solar value of [Z/H] \citep{LH05}, with the scatter imparting a $\lesssim$3\% error to the temperature fit for the majority of stars. The magnitude of this error is similar to that imparted by good-quality photometry.

A larger error is imparted by interstellar reddening, which makes the star appear dimmer and cooler than it actually is. This can be significant in the case of distant objects, or those in the Galactic Plane. In practice, stars which suffer from significant interstellar extinction tend to be the brighter giant stars, which also suffer from significant parallax errors. Bright giants are often subject to radial pulsations. These can change the fractional contribution of cool and hot spots on the stellar surface to the star's total light, leading to changes in the astrometric centre of light. These can impart substantial parallax errors \citep{vanLeeuwen07}, which can be sufficiently large that extinction is not the primary source of error in the placement of these stars on the H--R diagram.

As we do not know the mass of individual stars, we estimate it from the best-fit blackbody temperature and luminosity. As this only affects the stellar gravity, which has a minimal effect on the overall SED, we need only to approximate the actual mass. We estimate the stellar mass by assuming each star is either a main-sequence star or a giant. Giants are determined to be stars with:
\begin{equation}
	L > \left\{ 
		\begin{array}{rl}
			(2.25 \times 10^{-4}\ T)^7	&\mbox{ if $T>6000$ K} \\
			(6.5 \times 10^{-4}\ T)^7	&\mbox{ otherwise} \\
		\end{array} \right.
\end{equation}
where $T$ is the determined effective temperature in Kelvin and $L$ is the determined luminosity in solar units.

For main-sequence stars, we use a mass--temperature relation based on a solar-metallicity, zero-age main-sequence (ZAMS) isochrone \citep{DCJ+08}. We cannot estimate the mass of giant stars so easily. Most stars below the RGB tip ($L \approx 2500$ L$_\odot$) will be the more-numerous, older stars of $\sim$1 M$_\odot$. More massive giants survive to much higher luminosities on the AGB, thus we expect very luminous giants to be considerably more massive. Based on the aforementioned isochrones \citep{DCJ+08}, we assume a mass for the giant stars of $M = (L / 2500 {\rm L}_\odot)^{2/3} {\rm M}_\odot$, with limits placed at 1 and 20 M$_\odot$. 

The first-order determination of temperature, luminosity and stellar gravity (from the black-body fit), are used as the initial parameters for our synthetic spectra. Previously, we have used the {\sc marcs} model atmospheres \citep{GBEN75,GEE+08} described in \citet{MvLD+09}. For this work, however, we instead use the {\sc BT-Settl} models of \citet{AGL+03}, as the temperature spacing of the grid models is finer in the 4000--6000 K region. We have found this to reduce artifacts caused by interpolation between grid points for stars with poor-quality photometry.

{\sc Getsed} takes the model spectra grid and performs a linear interpolation in temperature and surface gravity (and metallicity, if required). The synthetic spectrum is convolved with the observed filter transmissions and reduced to a set of expected photometric fluxes. These are then normalised to the observed photometric fluxes (the constant of normalisation determining the luminosity) and a $\chi^2$ is determined.

This process is first performed on the temperature gridpoint immediately cooler than the blackbody temperature. {\sc Getsed} then calculates the $\chi^2$ for the neighbouring temperature gridpoints and continues until a $\chi^2$ minimum is detected. The temperature corresponding to the $\chi^2$ minimum is used as a new starting point, a new surface gravity is calculated, and $\chi^2$ is determined for 128 K steps between the neighbouring models. A new $\chi^2$ minimum is determined, the temperature step is halved, and the process re-run until the temperature is fit to within 1 K. The calculated stellar effective temperature, luminosity, surface gravity and model photometric fluxes are written to disk.

As with our previous uses of this code, we have only fit photometric data with short wavelengths (here we require $\lambda < 8$ $\mu$m). Circumstellar dust will still cause some opacity in the optical, but the obscuration must be relatively small for it to have been observed with \emph{Hipparcos}, and absorption of optical flux is relatively easy to identify when it is re-radiated in the infrared.

Stars which are heavily extincted will have SEDs that become double-peaked. In these cases, {\sc Getsed} will not be able to fit a model spectrum to it. The most-extincted stars (e.g.\ IRC+10216; \citealt{KH77}) may be sufficiently optically obscured that they do not feature in the \emph{Hipparcos} catalogue (cf.\ \citealt{BSvL+09}). This becomes important in the removal of bad data (see Section \ref{BadSect}) and we remind the reader that our H--R diagram is therefore incomplete, even at high luminosity.

\subsection{Removing bad data}
\label{BadSect}

Each star now has a series of photometric datapoints for which an observed and a modelled flux is known. The ratio ($R$) of observed/modelled flux therefore gives the excess or deficit flux in that band. We give this as a pure ratio, rather than in terms of an $\sigma$-based excess, as we do not include the errors in the photometric data in our model. While this may appear surprising, the reported errors on photometric data are almost invariably much lower than the absolute error between catalogues. Such `bad' data can be incorporated into the SED for several intrinsic and extrinsic reasons, including (in approximate order of overall severity):
\begin{list}{\labelitemi}{\leftmargin=1em \itemsep=0pt}
\item poor-quality raw data, e.g.\ saturated images or unflagged cosmic rays;
\item blending, particularly among catalogues which have integrated fluxes over different areas (e.g.\ the point-spread function of \emph{Hipparcos} is a very different size to that of \emph{IRAS}), affecting binary and multiple stars, and objects which are in the same line of sight as background objects with very red colours (e.g.\ redshifted galaxies in infrared surveys);
\item poor background subtraction, which mainly affects infrared observations of sources at low Galactic latitudes or in other regions of nebulosity;
\item intrinsic variability among stars;
\item inaccurate source matching due to large proper motions;
\item the accuracy of photometric corrections to each survey's base photometric system.
\end{list}
The issues become particularly problematic when comparing optical and mid-infrared catalogues: in the mid-infrared stars typically present fainter detections compared to higher backgrounds, and beam sizes are typically larger (hence include more objects). Additional errors come from the model parameters which arise from our assumption that stars have solar metallicity, a given stellar mass and no interstellar extinction. Bearing this in mind, we have assumed that each photometric point has an arbitrary absolute error of 10\% and computed a $\chi^2$-based goodness-of-fit measure based on this uncertainty. We also do not give error estimates for our temperatures and luminosities, as they would essentially be meaningless.

The bad data in our combined multi-wavelength catalogue is mostly of sources which have saturated in the various input survey catalogues. The \emph{WISE} catalogue, in particular, suffers from saturation. Some photometry flagged as good does not match detections in other bands (e.g.\ HIP\,24\,436 has $W_2$ = 1.987 $\pm$ 0.009 mag, while $W_{1,3,4}$ = 0.739, 0.001 and --0.030 mag). Conversely, some photometry flagged as uncertain does not decrease the goodness-of-fit of the model SEDs and is therefore sufficiently accurate for the purpose of identifying infrared excess. We have so far included all \emph{WISE} data, regardless of its uncertainty and we must now remove the points we believe to be in error.

Unfortunately, this is a particularly recalcitrant dataset to remove bad data from: we wish to keep points which fit badly due to intrinsic variability (as these will, on average, cancel out across the SED), but remove datapoints which have incorrect fluxes. We have opted to apply a number of sequential cuts to remove bad quality data. Since altering one band affects the model fit of the others, we must carefully design these cuts to minimise errant removal of good quality data. At each step, we have visually examined the SEDs of the objects with the worst-fitting models and devised a cut which removes the dominant contribution of bad data. The removed data was examined and the cut applied if it did not remove any plausibly accurate data. The details of these cuts are given in the Appendix.

While these cuts have not removed every single bad data point, they have removed the vast majority of bad data, providing a much cleaner data set with which we can work. This has sadly meant removing stars where there was not sufficient data to provide a robust fit, meaning the original \emph{Hipparcos} catalogue has been reduced from 117\,956 to 107\,619 objects.

\subsection{Defining infrared excess}
\label{IRXSSect}

Now that we have removed bad data from our catalogue, we can calculate the amount of infrared excess present for each star. Having already performed SED fitting, creating a measure of infrared excess becomes a trivial exercise in comparing the fitted model with observations. Providing the best metric(s) to quantify infrared excess is more difficult. We adopt two techniques.

Our first metric simply takes the ratio of observed to SED-modelled flux of all the data longward of 2.2 $\mu$m ($WISE$, $MSX$, $AKARI$ and $IRAS$), and averages them together. This provides a single number ($E_{\rm IR}$) that describes the average excess in the 3--25 $\mu$m region, relative to the underlying photospheric model, which can be described mathematically as:
\begin{equation}
E_{\rm IR} =
\sum_{\lambda > 2.2 \mu{\rm m}} \frac{F_\nu^{\rm obs} / F_\nu^{\rm model}}{n_{\rm obs}} ,
\label{EIREq}
\end{equation}
where $F_\nu^{\rm obs}$ and $F_\nu^{\rm model}$ are the observed and modelled fluxes at frequency $\nu$ (corresponding to wavelength $\lambda$) and $n_{\rm obs}$ is the number of observations at wavelengths $>$2.2 $\mu$m.

Our second metric assumes that the infrared excess is due to reprocessed stellar light (i.e.\ ignoring background infrared emission and foreground circumstellar extinction). We approximate the amount of reprocessed light as being the integrated observed flux\footnote{Here, the flux is defined in energy terms, i.e.\ $\int F_\nu {\rm d}\nu$ or $\int \lambda F_\lambda {\rm d}\lambda$.} longward of 2.2 $\mu$m, minus the integrated model flux over the same region. We can take this as a fraction of the underlying stellar flux, under the assumption that the total energy output (in Jy) of the star is not affected by circumstellar reprocessing of light. Mathematically, we can then define the fraction of stellar light reprocessed into the infrared ($L_{\rm IR} / L_\ast$) as:
\begin{equation}
\frac{L_{\rm IR}}{L_\ast} =
\frac{\int_{2.2 \mu{\rm m}}^{\infty} (F_\nu^{\rm obs} - F_\nu^{\rm model}) \  {\rm d}\nu}{\int_{0}^{\infty} F_\nu^{\rm obs} {\rm d}\nu} .
\label{LIREq}
\end{equation}
We also define a wavelength, $\lambda_{\rm peak}$, as where $\nu F_\nu^{\rm obs} - \nu F_\nu^{\rm model}$ reaches a peak. The precision with which we can define $\lambda_{\rm peak}$ depends strongly on the amount of available data.

\section{Master catalogue and H--R diagram}
\label{HRSect}

\subsection{Presenting the catalogue}
\label{HRDSect}

\begin{figure*}
 \resizebox{0.937\hsize}{!}{\includegraphics[angle=270]{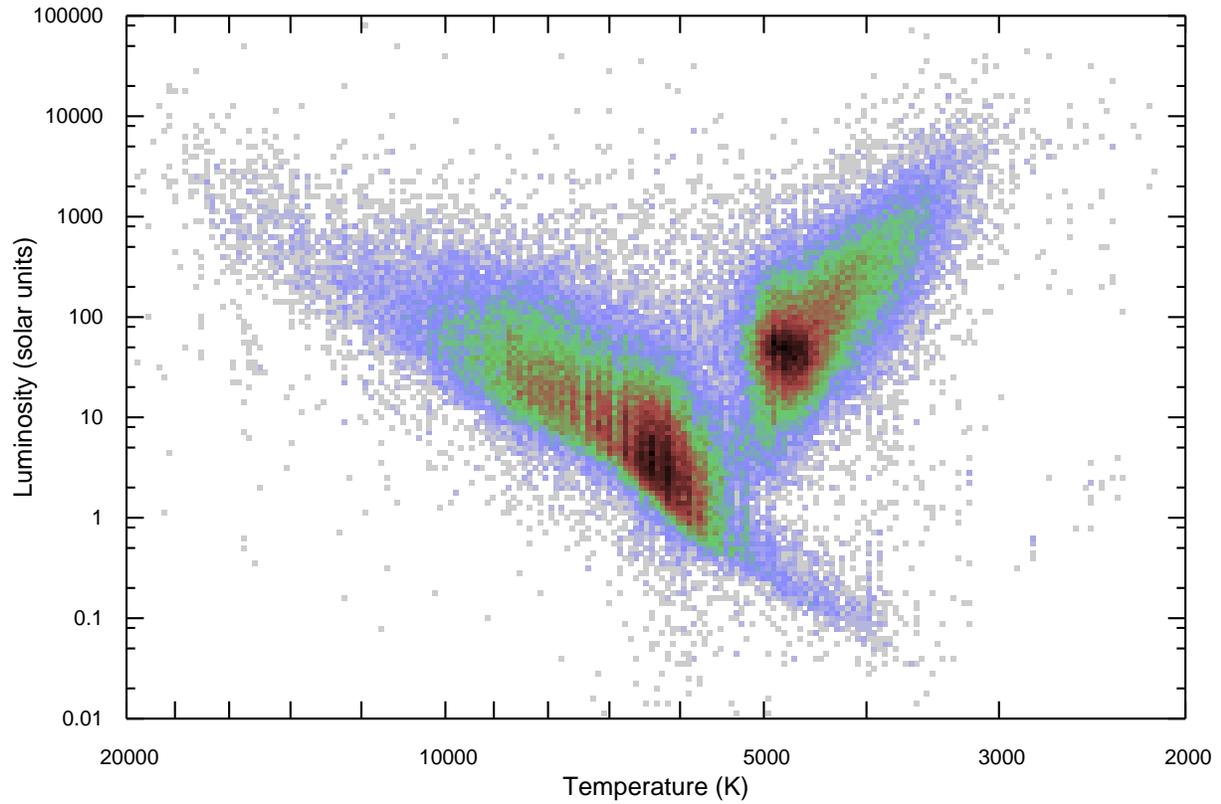}}
 \resizebox{0.937\hsize}{!}{\includegraphics[angle=270]{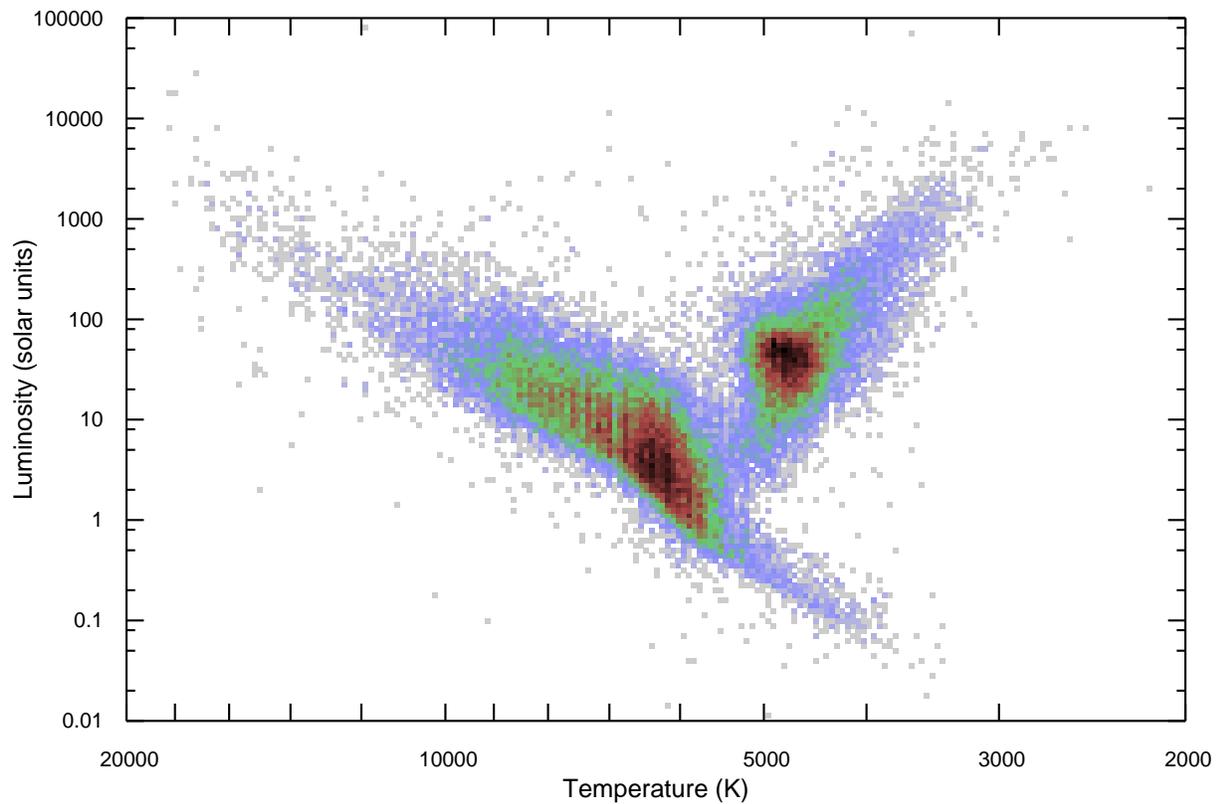}}
 \caption{Density-coded Hertzsprung--Russell diagram for the clipped \emph{Hipparcos} dataset, based on the {\sc BT-Settl} models. Top panel: stars with \emph{Hipparcos} distances of $<$1 kpc. Bottom panel: stars with \emph{Hipparcos} distances of $<$300 pc with parallax errors of $<$30\% and photometric data at $>$2.2 $\mu$m. Darker/redder colours show regions with a greater number of stars.}
 \label{HRDFig}
\end{figure*}

We are now in a position to list our data in a master catalogue containing the fundamental parameters of the \emph{Hipparcos} stars. We do so in Table \ref{HRDTable}, which lists:
\begin{list}{\labelitemi}{\leftmargin=1em \itemsep=0pt}
\item Column 1: the \emph{Hipparcos} identifier for the object;
\item Columns 2 \& 3: the co-ordainates of the object;
\item Columns 4 \& 5: the parallax distance to the object and its associated fractional error;
\item Column 6: the modelled effective temperature of the object;
\item Column 7: the modelled luminosity of the object;
\item Columns 8 \& 9: the shortest and longest wavelength for which we have data;
\item Columns 10--30: the ratio of the observed to modelled flux at each wavelength, such that unity represents a perfect match to the model;
\item Column 31: $n_{\rm IR}$, the number of observations at $\lambda > 2.2 \mu$m;
\item Column 32: $E_{\rm IR}$, the average excess at those wavelengths, as defined above;
\item Column 33: $L_{\rm IR} / L_\ast$, the fraction of the object's luminosity reprocessed into the infrared, as defined above;
\item Column 34: $\lambda_{\rm peak}$, the wavelength at which the infrared excess reaches a peak, as defined above.
\end{list}

\begin{center}
\begin{table*}
\caption{Fundamental parameters and infrared excess for \emph{Hipparcos} stars. The columns are described in the text. The complete table is available in the online version of the paper.}
\label{HRDTable}
\begin{tabular}{@{}r@{\ \ }r@{\ }r@{\ }r@{}r@{\ }r@{\ }r@{\ }r@{}r@{}r@{}r@{}r@{}r@{\ }r@{\ }r@{\ }r@{}}
    \hline \hline
\multicolumn{1}{c}{\clap{HIP}} & \multicolumn{1}{c}{RA} & \multicolumn{1}{c}{Dec} & \multicolumn{1}{c}{$d$} & \multicolumn{1}{c}{$\delta d$/$d$} & \multicolumn{1}{c}{$T_{\rm eff}$} & \multicolumn{1}{c}{$L$} & \multicolumn{2}{c}{\clap{Coverage (nm)}} & \multicolumn{1}{c}{SDSS $u$} & \multicolumn{1}{r}{\clap{\nodata}} & \multicolumn{1}{c}{\clap{IRAS [25]}} & \multicolumn{1}{c}{$n_{\rm IR}$} & \multicolumn{1}{c}{$E_{\rm IR}$} & \multicolumn{1}{c}{$\underline{L_{\rm IR}}$} & \multicolumn{1}{c}{$\lambda_{\rm peak}$}\\
\   & \multicolumn{1}{c}{\clap{(J2000)}} & \multicolumn{1}{c}{(J2000)} & \multicolumn{1}{c}{(pc)} & \multicolumn{1}{c}{} & \multicolumn{1}{c}{(K)} & \multicolumn{1}{c}{(L$_\odot$)} & \multicolumn{1}{r}{Start} & \multicolumn{1}{r}{End} & \multicolumn{1}{c}{Excess} & \multicolumn{1}{r}{\clap{\nodata}} & \multicolumn{1}{c}{Excess} & \multicolumn{1}{c}{\ } & \multicolumn{1}{c}{\ } & \multicolumn{1}{c}{$L_\ast$} & \multicolumn{1}{c}{\clap{($\mu$m)}}\\
    \hline
1 & 0.00091 & +01.08901 &  219.78 & 0.29 & 6400 &    8.73 & 354 &   2200 & 0.872 & \clap{\nodata} & 0.000 & 0 & 0.000 & 0 & 0 \\
2 & 0.00380 &--19.49884 &   47.96 & 0.05 & 3300 &    3.11 & 354 &    623 & 1.077 & \clap{\nodata} & 0.000 & 0 & 0.000 & 0 & 0 \\
3 & 0.00501 & +38.85929 &  442.48 & 0.15 & 8968 &  374.87 & 420 &   8610 & 0.997 & \clap{\nodata} & 0.000 & 1 & 2.300 & 0.0019 & 8.6 \\
\nodata&\nodata&\nodata&\nodata&\nodata&\nodata&\nodata&\nodata&\nodata&\nodata&\clap{\nodata}&\nodata&\nodata&\nodata&\nodata&\nodata\\
    \hline
\end{tabular}
\end{table*}
\end{center}

The catalogue is displayed as a Hertzsprung--Russell (H--R) diagram in Figure \ref{HRDFig}. The top panel of this Figure shows the \emph{Hipparcos} sample to separate out quite cleanly into the two traditional populations: the main sequence stars, which are largely complete for stars brighter than a few solar luminosities, and the giant branch stars, which lie to cooler temperatures. The concentration of stars on the giant branch is due to two groups of stars. The first being the horizontal branch stars (which, since the sample is largely metal-rich, form a red clump). The second being the RGB bump, the position which is also affected by metallicity \citep{CL02}. A significant scatter of stars is seen away from these two groups, which is not necessarily real and which we discuss in Section \ref{ResampleSect}.

\subsection{Distance-limiting the sample and associated biases}
\label{ResampleSect}

Despite the removal of a significant amount of bad data, there is still a large amount of scatter in the H--R diagram. There are four reasons for this. Firstly, a large number of \emph{Hipparcos} stars have relatively poor-quality data\footnote{We refer here to noise over the entire SED, rather than one or two clearly mismatching `bad' points.}. These are mostly stars where there is insufficient infrared data constraining the SED (these can be identified as those stars in Table \ref{HRDTable} where $n_{\rm IR}$, the number of measurements at $\lambda > 2.2$ $\mu$m, is small). This leads to a large fraction of the scatter observed in the H--R diagram, and to vertical artefacts (concentrations and rarefactions) on the {\sc BT-Settl} model grid spacing.

Secondly, stars which do have infrared excess or suffer from substantial interstellar extinction scatter towards cooler temperatures (and, in the case of interstellar extinction, lower luminosities) in the H--R diagram, as their optical light is either reprocessed into the infrared or scattered out of the line of sight.

Thirdly, and perhaps most importantly, scatter arises from uncertainty in the \emph{Hipparcos} parallax, which smears objects vertically in the H--R diagram. This also leads to the Lutz--Kelker bias \citep{LK73}. This bias occurs due to the inversion of parallactic angle to obtain a distance. As the measurement error is in parallax, this preferentially scatters objects to smaller distances. \emph{Hipparcos} data suffers from this significantly. We use the benchmark of 17.5\% error in parallax (which corresponds to an average 30\% deviation in a set of stellar luminosities; \citealt{OGS98}) as our figure of merit. Of the 107\,619 stars in our final sample, only 49\,188 have parallax errors less than this value. This rather severe limitation reduces the usefulness of the sample in examining stellar populations, particularly for the relatively-rare stars on the upper giant branches. For many applications a wider sample, with increased Lutz--Kelker bias is preferable. We therefore continue to include objects susceptable to significant Lutz--Kelker bias, but warn the reader to be mindful of its existence.

Finally, an additional distance error is present in red giants, where changes in brightness across the stellar surface (which covers a finite solid angle) cause a measurable astrometric shift. This can be misinterpreted as a parallactic shift, leading to much smaller \emph{Hipparcos} distances than their true distance. This is perhaps best observed in the case of W Lyn, which has a \emph{Hipparcos} parallax of 21.53 $\pm$ 8.06 mas, despite being several kpc distant \citep{IDF+01}. This is an extreme case, though we warn the reader that no supposedly volume-limited sample of any consequence will be clean of all intruding objects for these reasons.

For the remainder of our analysis we adopt two volume-limited samples, which are subject to these biases at differing levels. The first is limited to stars with parallax distances of $<$300 pc with $<$30\% parallax errors, which have a wide range of data which cover the SED well ($\lambda_{\rm max} > 2200 \mu$m). This subset of data still contains 46\,869 of the original \emph{Hipparcos} stars, of which 34\,660 have parallax errors below 17.5\%. The second sample is distance limited to $<$200 pc with $<$30\% parallax errors, with the same requirement that in which 32\,741 out of 33\,898 stars have errors $<$17.5\% in parallax. We refer to these in the discussion as the 300-pc and 200-pc samples.


\section{Discussion}
\label{DiscSect}

\subsection{Stellar isochrones}
\label{IsoSect}

\begin{figure*}
 \resizebox{0.95\hsize}{!}{\includegraphics[angle=270]{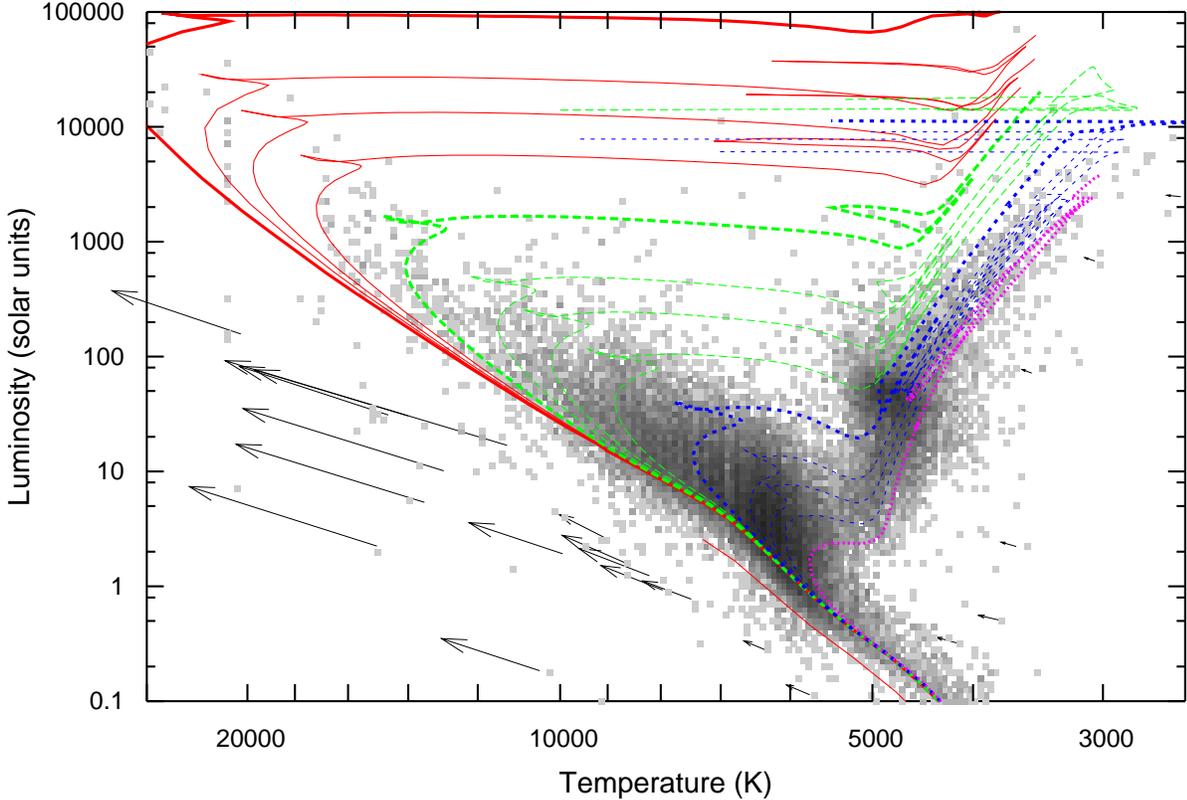}}
 \caption{Density-coded Hertzsprung--Russell diagram for the 200-pc sample (greyscale). Overplotted are solar-metallicity isochrones from the Padova models \protect\citep{MGB+08,BGMN08} at 10, 20, 30 and 50 Myr (solid, red lines); 100, 200, 300 and 500 Myr (long-dashed green lines); 1, 2, 3 and 5 Gyr (short-dashed blue lines); and 10 Gyr (dotted magenta line). The thin red line to the left of the main sequence is a zero-age isochrone at [Fe/H] = --1 to illustrate the blueward shift caused by decreasing metallicity. Black arrows show the effect of dereddening individual sources by $E(B-V) = 0.1$ mag.}
 \label{IsoHRDFig}
\end{figure*}

Much can be made of the H--R diagram in terms of the star formation history of the local neighbourhood. However, to do so thoroughly requires a more in-depth analysis than we are able to provide in this work. As a cursory analysis, we present solar-metallicity Padova isochrones \citep{MGB+08,BGMN08} for a variety of ages in Figure \ref{IsoHRDFig}. The H--R diagram is well-described by a population of mixed age, but of near-solar metallicity.

Interstellar reddening does not appear to be a great cause of concern in the 200-pc sample in general, with the majority of stars lying within the bounds of the isochrones with only a few tenths of a magnitude of dereddening at most. We show on Figure \ref{IsoHRDFig} the effect that a reddening of $E(B-V) = 0.1$ mag has on particular sources chosen at a variety of different temperatures. We can see here that cooler sources are largely unaffected by this modest reddening, but that the effect becomes much more severe as the peak of the SED becomes bluer and the short-wavelength photometry available fails to constrain the SED. This may lead to underestimates of the temperatures and luminosities of some of the hotter stars. Particularly, errors can be large if short-wavelength data is unavailable or unusable.

\subsection{Comparison to spectroscopic temperatures}
\label{SpecSect}

\begin{figure}
 \resizebox{0.95\hsize}{!}{\includegraphics[angle=270]{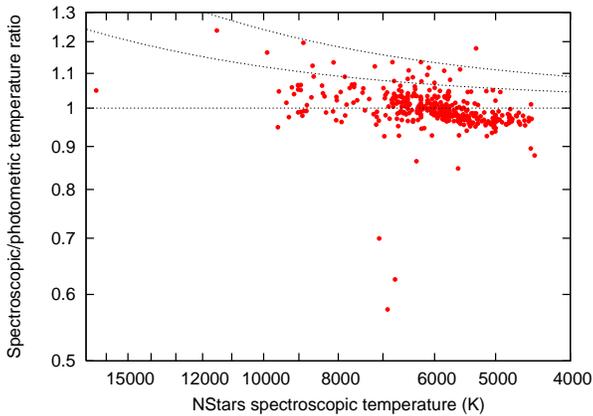}}
 \caption{Comparison of our temperatures with those derived spectroscopically from the NStars project. Lines show the approximate deviations expected for $E(B-V) \approx 0.0$, 0.1 and 0.2 mag.}
 \label{NstarsFig}
\end{figure}

With this in mind, we can check the consistency of our results, by comparing them to the spectroscopically-derived temperatures of the NStars project \citep{GCG+03,GCG+06}. This project identifies the basic parameters of stars within 40 pc of the Sun by fitting moderately-high-resolution spectra, thus the results are unbiased by interstellar reddening. Figure \ref{NstarsFig} shows the ratio of their temperatures to ours for 407 stars we have in common. Examination of the three outliers (HIP 35\,550, 59\,199 and 71\,957) in this figure show obvious problems with the 2MASS photometry that were missed by our bad data cuts. Neglecting these, the average temperature is consistent to $<$0.22\% (i.e.\ the error on the mean). The standard deviation of results is 4.4\% overall and decreases slightly toward lower temperatures.

Interstellar reddening should not affect the spectroscopic temperature, but imparts an apparent cooling to the photometric temperature, scattering points above unity. Four stars have a spectroscopic temperature $>$15\% higher than the photometric temperature: HIP 71\,193, 84\,379 ($\delta$ Her), 93\,805 ($\lambda$ Aql) and 98\,495 ($\epsilon$ Pav). The latter two scatter in this direction due to excess flux in the \emph{WISE} 4.6-$\mu$m band. The former two have SEDs that are poorly constrained by the input photometry. Unsurprisingly, we therefore find negligible reddening among stars within 40 pc.

In principle, with sufficiently-good-quality photometry, one could compute a three-dimensional extinction map of the Solar Neighbourhood by comparing spectroscopically-derived temperatures to photometrically-derived temperatures. The lack of self-consistent photometry for the \emph{Hipparcos} stars probably prevents such a determination here, but may become possible in the \emph{Gaia} era.

\subsection{Grouping objects by type}
\label{ItaSect}

\begin{figure*}
 \resizebox{0.95\hsize}{!}{\includegraphics[angle=270]{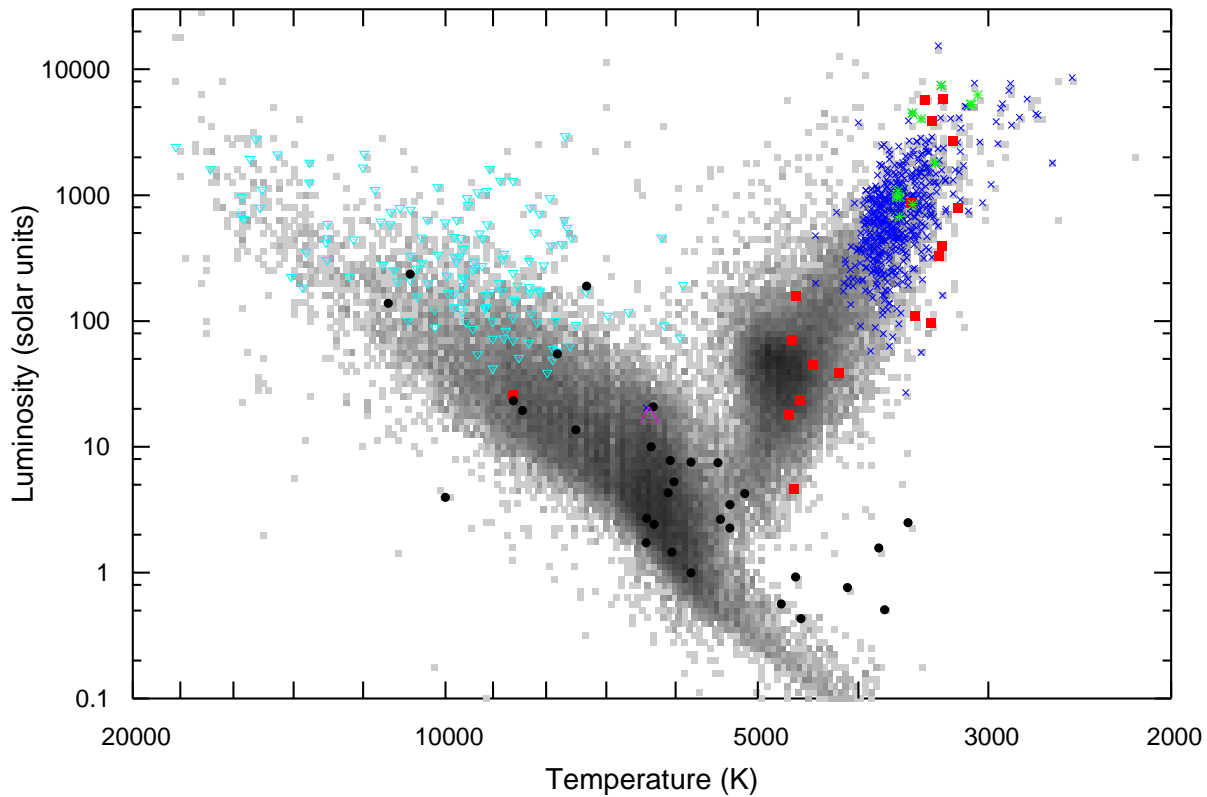}}
 \caption{Density-coded Hertzsprung--Russell diagram for the 300-pc sample (greyscale, from bottom panel of Figure \protect\ref{HRDFig}). The overplotted symbols are from \protect\citet{IMI+10} and show known examples of the following. Black filled circles: pre-main-sequence stars --- cyan downward-pointing triangles: Be stars --- large, magenta upward-pointing triangle: Wolf--Rayet star --- blue crosses: M-type (super-)giants --- green asterisks: S-type giants --- red filled squares: carbon stars.}
 \label{ItaHRDFig}
\end{figure*}
\begin{figure}
 \resizebox{0.95\hsize}{!}{\includegraphics[angle=270]{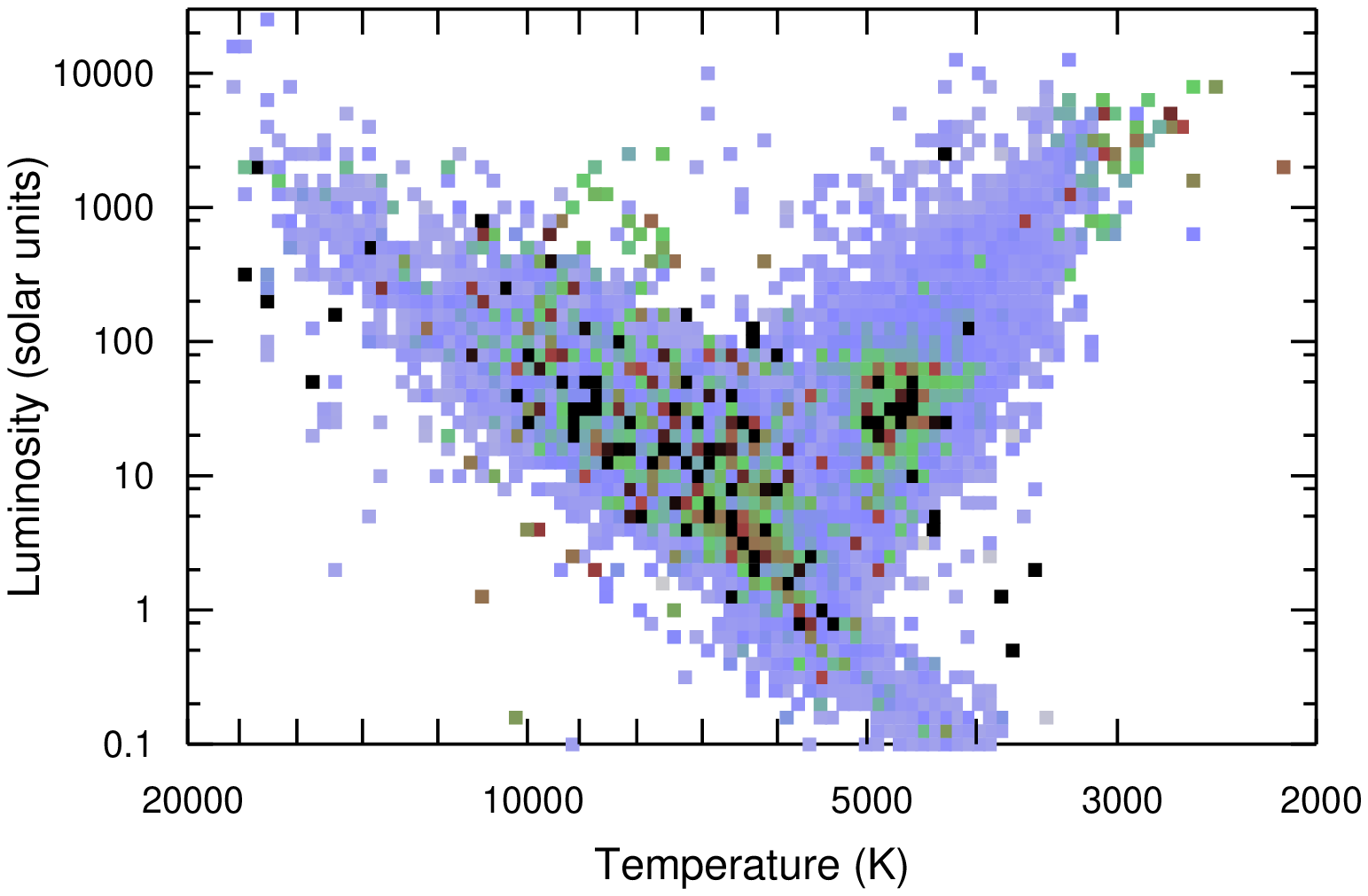}}
 \resizebox{0.95\hsize}{!}{\includegraphics[angle=270]{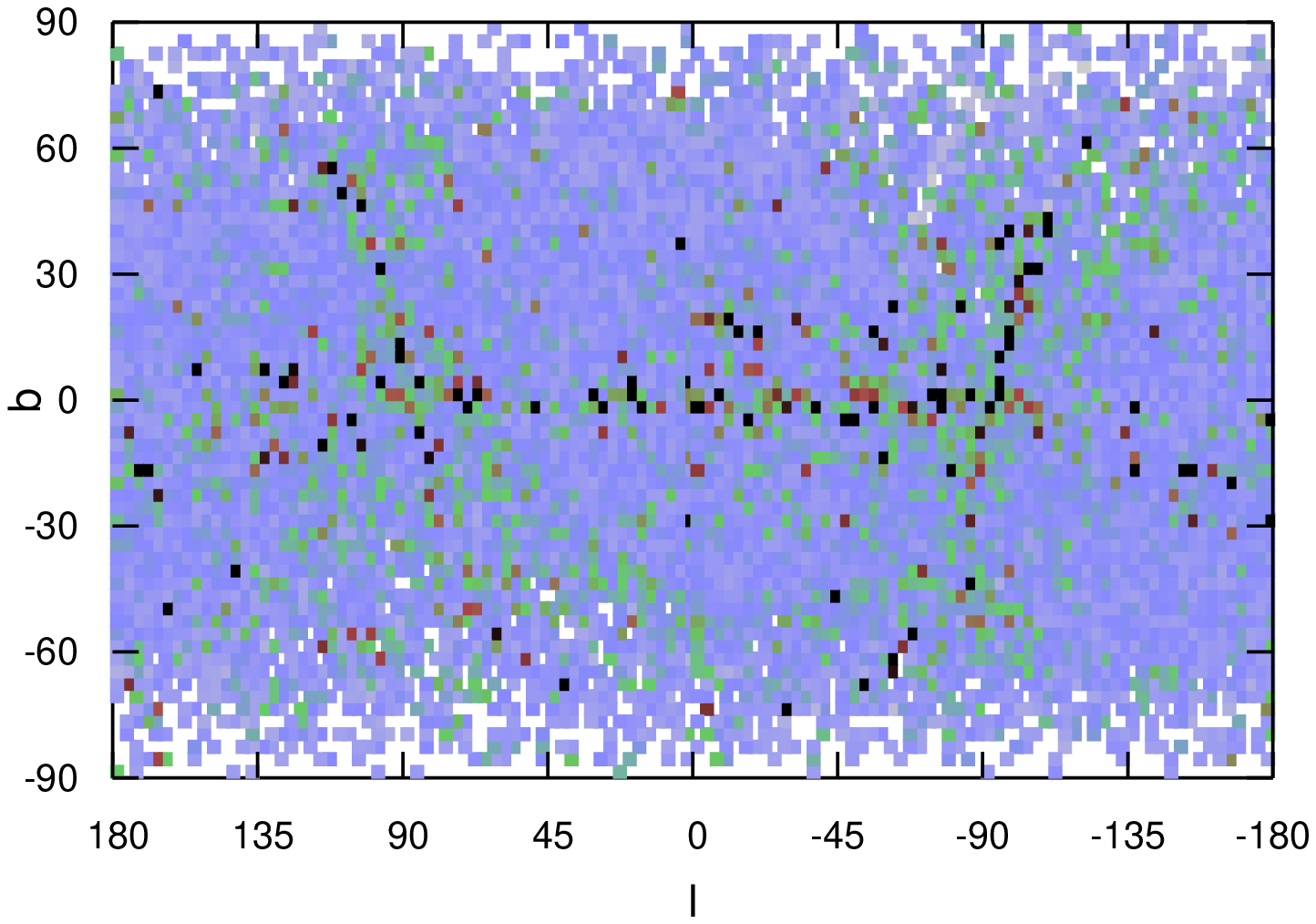}}
 \resizebox{0.95\hsize}{!}{\includegraphics[angle=270]{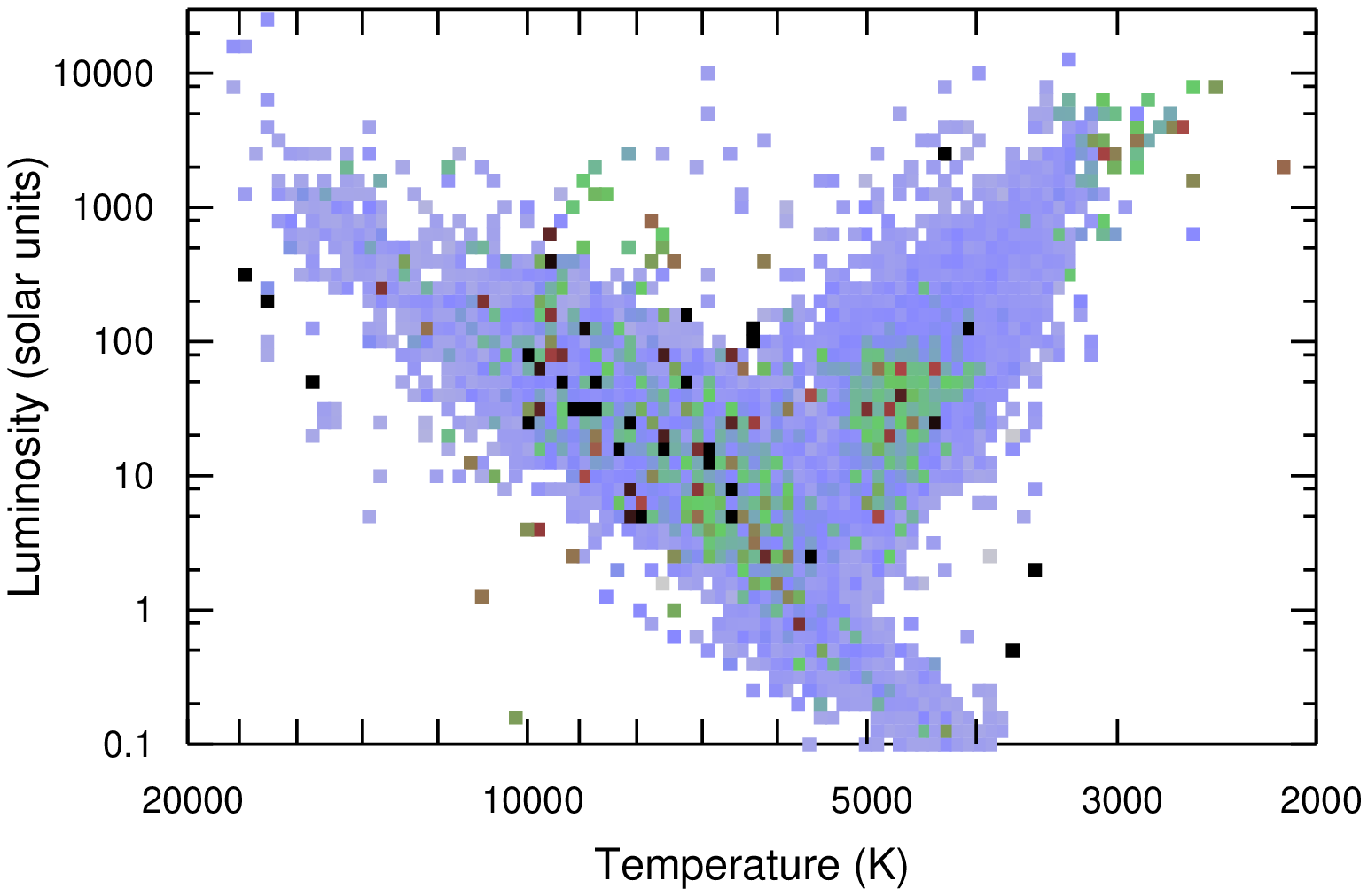}}
 \caption{Top panel: Hertzsprung--Russell diagram of the 300-pc \emph{Hippacros} sample. Bins are colour-coded by infrared excess ($E_{\rm IR}$), with the colour-coding denoting the object with the greatest excess in that bin: light blue indicates no excess ($E_{\rm IR} = 1$), green indicates moderate excess ($E_{\rm IR} \approx 2$), red strong excess ($E_{\rm IR} \approx 5$) and black extreme excess (capped at $E_{\rm IR} = 10$). Middle panel: as top panel, but showing the Galactic distribution of those sources. Bottom panel: as top panel, removing sources with $|b| < 5^\circ$ and within 5$^\circ$ of the \emph{IRAS} missing strip.}
 \label{XSHRDFig}
\end{figure}
\begin{figure}
 \resizebox{0.95\hsize}{!}{\includegraphics[angle=270]{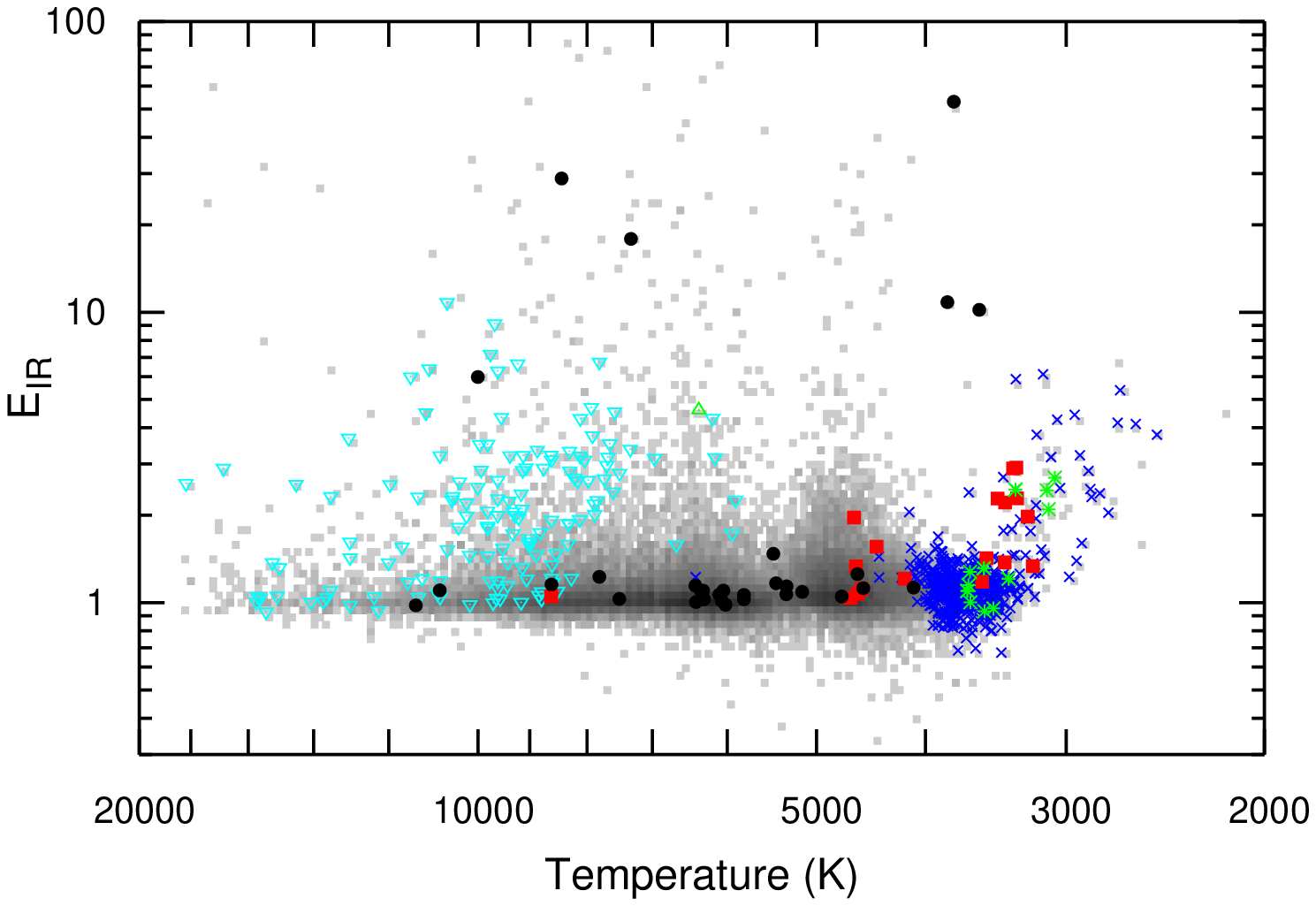}}
 \resizebox{0.95\hsize}{!}{\includegraphics[angle=270]{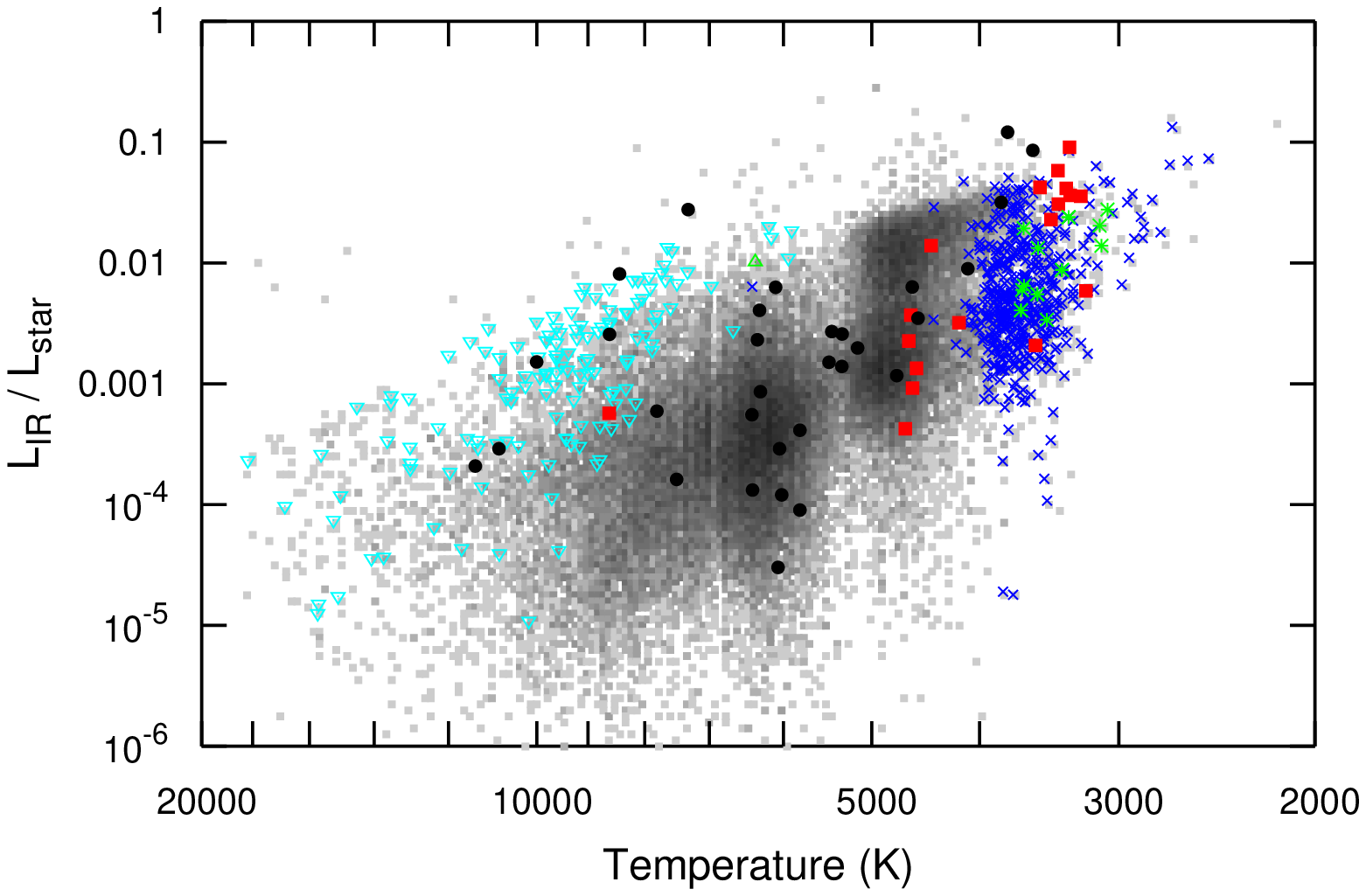}}
 \caption{Measures of excess for different kinds of identified stars (see text for definitions of $E_{\rm IR}$ and $L_{\rm IR}$). Symbols are as in Figure \protect\ref{ItaHRDFig}.}
 \label{ItaXSFig}
\end{figure}
\begin{figure}
 \resizebox{0.95\hsize}{!}{\includegraphics[angle=270]{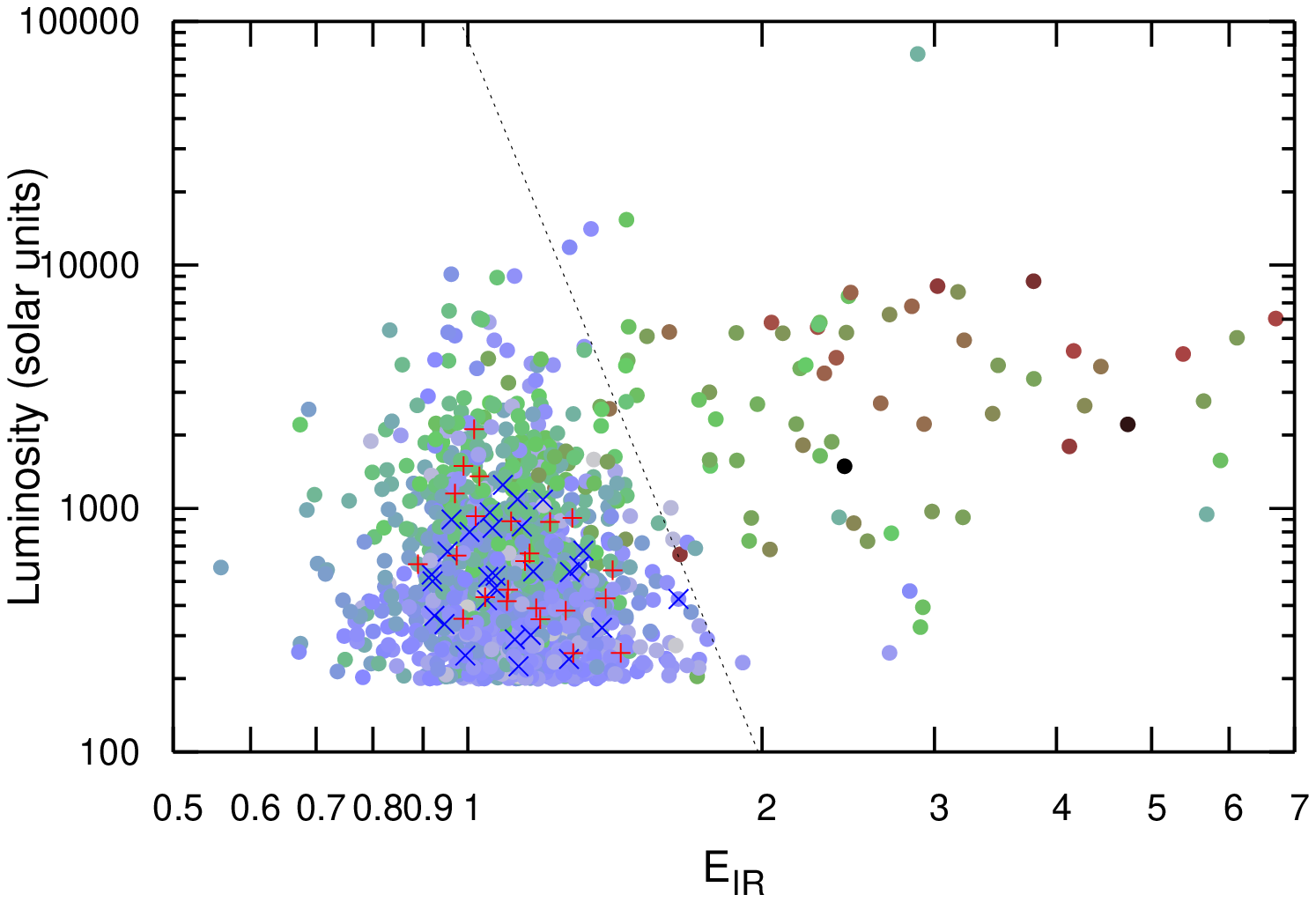}}
 \caption{Excess among giant stars. Colour scale denotes modelled stellar effective temperature: red points are coolest, blue points are warmest. Red plus signs show stars from \protect\cite{Groenewegen12} that were identified to have optical depths of $\tau_{\rm V} > 10^{-5}$, blue crosses show those stars that were not. The dotted black line shows our definition of those stars with infrared excess.}
 \label{GiantXSFig}
\end{figure}

The \emph{AKARI--Hipparcos} catalogue of \citet{IMI+10} measures excess at 9 and 18 $\mu$m, and contains grouping information for 2787 stars commonly exhibiting infrared excess, including carbon stars, red giants, supergiants, S-type stars, etc. \emph{AKARI} only detected 44\% of the \emph{Hipparcos} stars, so can be regarded as a selected subset of the \emph{Hipparcos} sample, subject to its own biases. We remind the reader that the \emph{Hipparcos} sample does not include all of the local ($<$300 pc) optically-obscured giant stars, and that the compilation of stellar types listed in Ita et al.\ are only as complete as the literature from which they are based. The catalogue from \citet{IMI+10} cannot therefore be treated as a definitive, complete list of each type of source, nor does it purport to be such. Of the 2787 classified stars in the list of Ita et al., we retain 2764 after removing bad data. Of these, only 749 meet the criteria for our 300-pc sample and 293 for our 200-pc sample.

The objects classified in \citet{IMI+10} within the 300-pc sample are presented on a H--R diagram in Figure \ref{ItaHRDFig}. In general, the different types of stars match up well with their expected locations. M giants and S stars lie on the upper giant branch, along with the majority of carbon stars (we have not made an effort here to separate intrinsic from extrinsic carbon stars \citep{VEJU+98} due to the incomplete nature of any determination). Be stars (which include a variety of hot, emission-line stars) are located in the upper-left of the H--R diagram, but scatter toward temperatures cooler than the main sequence due to reprocessing of stellar light into the infrared by the circumstellar excretion disc (cf.\ \citealt{KBSF06}). Pre-main-sequence (pre-MS) stars likewise mostly lie toward the cooler side of the main sequence due to reprocessing of their optical emission into the infrared, lowering the effective temperature.

The top panel of Figure \ref{XSHRDFig} shows a similar H--R diagram, indicating the location of the stars with greatest infrared excess, as measured in terms of average fractional excess over the mid-infrared spectrum ($E_{\rm IR}$). The two regions of high stellar density (the main sequence above 6000 K, and the red clump) contain the largest number of infrared-excessive stars. These stars are not usually truly excessive, except the case of several pre-MS stars, but are instead artifically reddened or suffer from source confusion in the infrared. The majority of these stars are located near the Galactic Plane, or lie in or near the \emph{IRAS} missing stripe and thus suffer from poor infrared photometry. Some stars in the Plane may suffer from sufficient interstellar extinction to appear to have infrared excess, even at distances as small as 300 pc. Removing sources within 5$^\circ$ of the Galactic Plane and the \emph{IRAS} missing strip yields the H--R diagram at the bottom of Figure \ref{XSHRDFig}. The number of highly-excessive sources is greatly reduced, with the few remaining sources of high excess located predominantly in the Gould Belt.

More pertinently, two further groups of infrared-excessive stars in Figure \ref{XSHRDFig} are largely unchanged by this process. The first, above the main sequence at around 10\,000 K, are identified as Be/Ae stars by \citet{IMI+10}; the second are the cool, luminous stars near to and above the RGB tip. This second group are identified as M giants and S and carbon stars by \citet{IMI+10}, and are likely to entirely be dust-producing AGB stars.

In Figure \ref{ItaXSFig}, we show the different types of identified stars from \citet{IMI+10} as a function of our two measures of infrared excess: $E_{\rm IR}$ and $L_{\rm IR}/L_\ast$. While $L_{\rm IR}/L_\ast$ is a more physical measure of infrared excess, $E_{\rm IR}$ is clearly more effective at separating out the infrared excessive stars from the bulk of the population, particularly for the Be stars.

Figure \ref{ItaXSFig} also shows a signficiant scatter below $E_{\rm IR} = 1$. This represents a supposed deficit of infrared flux in these stars, and tends to be much more prevelant in the cooler stars. This is largely due to decreased sensitivity in stars further down the main sequence, but some upper-giant-branch stars also have infrared deficits. In this case, scatter can be introduced by stellar variability, as photometry is not usually averaged or taken contemporaneously, and does not imply either an instantaneous or time-averaged deficit in infrared flux.

\subsection{Giant stars with excess}
\label{GiantXSSect}

\begin{center}
\begin{table*}
\caption{Literature spectroscopic and variability information for luminous ($>$850 L$_\odot$) giant stars with detected circumstellar emission. Details of columns and explanations of acronymns are listed in the text.}
\label{GiantsTable}
\begin{tabular}{@{}r@{\quad}l@{\quad}l@{\ }r@{\ }r@{\quad} l@{\ }l@{\ }c@{\ }c@{\ }r@{\ }c r@{\ }r@{\quad}r@{}}
    \hline \hline
HIP	& Name	& Spectral 	& Temper-	& Lumin-	&\multicolumn{6}{c}{Variability}	& LRS	& SWS & $E_{\rm IR}$\\
\ 	& \ 	& Type		& ature		& osity		&Type & Amplitude & Band & Source & Period & Source & Classif-	& Classif- & \ \\
\ 	& \ 	& \ 		& (K) 		& (L$_\odot$)	& \ & (mag) & \ & \ & (days) & & ication & ication & \ \\
    \hline
1728   & T Cet           & M5-6SIIe   & 3329 &  7442 & SRV  & 1.9  & V & G & 159     & G & \nodata & \nodata & 2.45 \\
8565   & TT Per          & M5II-III   & 3228 &  1579 & SRV  & 1.4  & p & G & 82      & G & SE7     & \nodata & 1.89 \\
9234   & V370 And        & M7III      & 2948 &  3831 & SRV  & 1.01 & H & G & 228     & G & \nodata & \nodata & 4.44 \\
13064  & Z Eri           & M5III      & 3354 &  2334 & SRV  & 1.63 & V & G & 80      & G & SE8     & \nodata & 1.79 \\
17881  & SS Cep          & M5III      & 3158 &  5104 & SRV  & 1.1  & p & G & 90      & G & SE3     & \nodata & 1.53 \\
24169  & RX Lep          & M7         & 3256 &  3764 & SRV  & 2.4  & V & G & 60      & G & SE6     & \nodata & 2.19 \\
25194  & SW Col          & M1III      & 3661 &   921 & V    & 0.34 & V & G & \nodata & \nodata & SE8     & \nodata & 2.40 \\
27989  & $\alpha$ Ori    & M2Iab:     & 3659 & 73524 & SRV  & 1.3  & V & G & 2335    & G & \nodata & 2.SEcp  & 2.88 \\
28166  & BQ Ori          & M5IIIv     & 3192 &   917 & SRV  & 2.1  & V & G & 110     & G & \nodata & \nodata & 1.95 \\
28816  & SS Lep          & A1V+M6II   & 4347 &  2672 & Bin  & 0.24 & V & G & \nodata & \nodata & \nodata & \nodata & 12.60 \\
28874  & S Lep           & M5III      & 3187 &  3415 & SRV  & 1.58 & V & G & 89      & G & SE6t    & \nodata & 3.79 \\
36288  & Y Lyn           & M6SIb-II   & 3110 &  5249 & SRV  & 2.5  & V & G & 110     & G & \nodata & \nodata & 2.10 \\
38834  & V341 Car        & M0III      & 3326 &  1580 & V    & 0.9  & V & G & 0       & G & \nodata & \nodata & 5.88 \\
41037  & $\epsilon$ Car  & K3:IIIv+?  & 4209 & 14086 & Bin  & 0.08 & H & G & 0       & G & \nodata & \nodata & 1.34 \\
42489  & RV Hya          & M5II       & 3200 &  1884 & SRV  & 1.3  & V & G & 116     & G & \nodata & \nodata & 2.36 \\
43215  & AK Pyx          & M5III      & 3410 &  1499 & V    & 0.42 & H & G & 0       & G & \nodata & \nodata & 1.77 \\
44050  & RT Cnc          & M5III      & 3192 &  2225 & SRV  & 1.48 & V & G & 60      & G & SE3     & \nodata & 2.17 \\
44862  & CW Cnc          & M6         & 2909 &  2228 & V    & 1.2  & p & G & \nodata & \nodata & SE3     & \nodata & 2.93 \\
45058  & RS Cnc          & M6IIIase   & 3122 &  5282 & SRV  & 1.5  & p & G & 120     & G & \nodata & \nodata & 2.44 \\
46806  & R Car           & M6.5IIIpev & 2800 &  4164 & Mira & 6.6  & V & G & 309     & G & SE1     & \nodata & 2.38 \\
48036  & R Leo           & M8IIIe     & 1995 &  1493 & Mira & 6.9  & V & G & 310     & G & SE2     & \nodata & 2.43 \\
51821  & U Ant           & C5,3(Nb)   & 3317 &  5819 & V    & 0.9  & p & G & \nodata & \nodata & SiC+: & \nodata & 2.29 \\
52009  & U Hya           & C6.5,3(N2) & 3400 &  3893 & SRV  & 2.4  & B & G & 450     & G & SiC & \nodata & 2.22 \\
53809  & R Crt           & M7III      & 2491 &  8591 & SRV  & 1.4  & p & G & 160     & G & SE3t    & \nodata & 3.79 \\
57607  & V919 Cen        & M7III      & 3094 &  7766 & SRV  & 0.58 & H & G & \nodata & \nodata & \nodata & \nodata & 3.17 \\
61022  & BK Vir          & M7III:     & 2889 &  2706 & SRV  & 1.52 & V & G & 150     & G & SE4t    & \nodata & 2.64 \\
63642  & RT Vir          & M8III      & 2602 &  1804 & SRV  & 1.29 & V & G & 155     & G & SE3t    & 2.SEa   & 4.12 \\
64569  & SW Vir          & M7III      & 2918 &  4917 & SRV  & 1.5  & V & G & 150     & G & SE3t    & \nodata & 3.22 \\
68357  & RW CVn          & M7III:     & 3141 &   973 & SRV  & 1.1  & p & G & 100     & G & SE2:    & \nodata & 2.98 \\
68815  & $\theta$ Aps    & M6.5III:   & 3151 &  3879 & SRV  & 2.2  & p & G & 119     & G & SE5t    & 2.SEb   & 3.48 \\
69816  & U UMi           & M6e        & 3018 &  1821 & Mira & 5.9  & V & G & 331     & G & SE2     & \nodata & 2.20 \\
70401  & RX Boo          & M7.5       & 2581 &  8196 & SRV  & 2.67 & V & G & 162     & G & SE3t    & 2.SEa   & 3.02 \\
70969  & Y Cen           & M7III      & 2907 &  5317 & Irr  & 1.1  & p & G & 180     & G & SE1t    & \nodata & 1.61 \\
71802  & RW Boo          & M5III:     & 3148 &  3010 & SRV  & 1.5  & V & G & 209     & G & SE7     & 2.SEb   & 1.77 \\
72208  & EK Boo          & M5III      & 3333 &  5587 & SRV  & 0.38 & H & G & \nodata & \nodata & \nodata & \nodata & 1.46 \\
76423  & $\tau_4$ Ser    & M5II-III   & 3165 &  5264 & SRV  & 1.18 & V & G & 100     & G & SE4     & \nodata & 1.88 \\
77619  & ST Her          & M6-7IIIaS  & 3071 &  6270 & SRV  & 1.5  & V & G & 148     & G & SE1     & 2.SEa   & 2.70 \\
78574  & X Her           & M8         & 3152 &  2765 & SRV  & 1.1  & V & G & 95      & G & SE6t    & 2.SEb   & 5.65 \\
80488  & U Her           & M7III      & 2700 &  4438 & Mira & 7    & V & G & 406     & G & SE4     & 2.SEc   & 4.16 \\
80704  & g Her           & M6III      & 3261 &  4056 & SRV  & 2    & V & G & 89      & G & \nodata & \nodata & 1.46 \\
84345  & $\alpha$ Her    & M5Iab:     & 3351 & 15368 & Bin  & 1.26 & V & G & \nodata & \nodata & \nodata & 1.NOp   & 1.45 \\
86527  & BM Sco          & K2.5Iab:   & 3676 &   949 & SRV  & 1.9  & p & G & 815     & G & \nodata & \nodata & 5.69 \\
94162  & SZ Dra          & M          & 3173 &   923 & V    & 1    & p & G & \nodata & \nodata & SE5     & \nodata & 3.21 \\
95413  & CH Cyg          & M7IIIv     & 2687 &  4316 & Bin  & 2.89 & V & G & \nodata & \nodata & \nodata & 2.SEc   & 5.39 \\
95902  & AF Cyg          & M4         & 3305 &  1646 & SRV  & 2    & V & G & 93      & G & SE3t    & \nodata & 2.29 \\
98031  & S Pav           & M7IIe      & 2752 &  5563 & SRV  & 3.8  & V & G & 381     & G & SE2t    & 2.SEa   & 2.28 \\
99082  & V1943 Sgr       & M7III      & 2752 &  5813 & V    & 2    & p & G & \nodata & \nodata & SE2t    & 2.SEa   & 2.04 \\
99990  & RT Cap          & C6,4(N3)   & 3245 &  2683 & SRV  & 2.8  & p & G & 393     & G & SiC+: & \nodata & 1.98 \\
100935 & T Mic           & M7III      & 2856 &  7708 & SRV  & 1.9  & p & G & 347     & G & SE1t    & 2.SEa   & 2.46 \\
101810 & EU Del          & M6III      & 3227 &  1585 & SRV  & 1.11 & V & G & 60      & G & N       & \nodata & 1.77 \\
104451 & T Cep           & M5-9       & 2866 &  6767 & Mira & 6.1  & V & G & 388     & G & SE1     & 2.SEa   & 2.85 \\
107516 & EP Aqr          & M8IIIv     & 3056 &  2651 & SRV  & 0.45 & V & G & 55      & G & SE5t    & 2.SEb   & 4.27 \\
108928 & TW Peg          & M7.5IIIv   & 3145 &  5027 & SRV  & 0.9  & p & G & 929     & G & SE6t    & \nodata & 6.11 \\
110396 & DZ Aqr          & M          & 3055 &  2454 & V    & 1.1  & V & G & \nodata & \nodata & \nodata & 2.SEb   & 3.44 \\
110428 & BW Oct          & M7III      & 2849 &  3592 & V    & 0.9  & p & G & \nodata & \nodata & SE5t    & \nodata & 2.32 \\
114318 & Y Scl           & M6III      & 3039 &   872 & SRV  & 1.6  & p & G & \nodata & \nodata & SE7     & \nodata & 2.48 \\
114404 & V345 Peg        & M3         & 3345 &  2795 & V    & 0.37 & H & G & \nodata & \nodata & \nodata & \nodata & 1.72 \\
117245 & TX Psc          & C7,2(N0)   & 3451 &  5693 & V    & 0.41 & V & G & \nodata & \nodata & N & 1.NC & 2.28 \\
118188 & R Cas           & M7IIIe     & 2187 &  2219 & Mira & 8.8  & V & G & 431     & G & SE5t    & 2.SEb   & 4.73 \\
    \hline
\end{tabular}
\end{table*}
\end{center}

\begin{center}
\begin{table*}
\caption{Literature spectroscopic and variability information for less-luminous ($<$1000 L$_\odot$) giant stars with detected circumstellar emission. Details of columns and explanations of acronymns are listed in the text.}
\label{Giants2Table}
\begin{tabular}{@{}r@{\quad}l@{\quad}l@{\ }r@{\ }r@{\quad} l@{\ }l@{\ }c@{\ }c@{\ }r@{\ }c r@{\ }r@{\quad}r@{}}
    \hline \hline
HIP	& Name	& Spectral 	& Temper-	& Lumin-	&\multicolumn{6}{c}{Variability}	& LRS	& SWS & $E_{\rm IR}$\\
\ 	& \ 	& Type		& ature		& osity		&Type & Amplitude & Band & Source & Period & Source & Classif-	& Classif- & \ \\
\ 	& \ 	& \ 		& (K) 		& (L$_\odot$)	& \ & (mag) & \ & \ & (days) & & ication & ication & \ \\
    \hline
893    & AC Cet          & M3III      & 3413 &   793 & SRV  & 0.33 & V & G & \nodata & \nodata & \nodata & \nodata & 2.71 \\
39751  & RU Pup          & C5,4(N3)   & 3323 &   394 & SRV  & 1.9  & p & G & 425     & G & \nodata & \nodata & 2.92 \\
43438  & RS Cam          & M4III      & 3298 &   739 & SRV  & 1.8  & V & G & 89      & G & \nodata & \nodata & 1.94 \\
50916  & HR 4091         & K4III      & 4057 &   459 & V    & 0.03 & H & H & \nodata & \nodata & \nodata & \nodata & 2.83 \\
52656  & TZ Car          & C (R5)     & 3346 &   326 & SRV  & 1.7  & p & G & 69      & G & \nodata & \nodata & 2.90 \\
57800  & RU Crt          & M3         & 3054 &   681 & V    & 1    & p & G & \nodata & G & SE3:    & \nodata & 2.04 \\
59389  & HD 105822       & K0/K1III   & 4464 &   234 & None & 0.03 & H & H & \nodata & \nodata & \nodata & \nodata & 1.91 \\
59458  & 68 UMa          & K5III      & 4478 &   256 & None & 0.04 & H & H & \nodata & \nodata & \nodata & \nodata & 2.70 \\
71568  & HR 5464         & K4III      & 4214 &   727 & None & 0.05 & H & H & \nodata & \nodata & \nodata & \nodata & 1.68 \\
112155 & BD Peg          & M8         & 3147 &   736 & SRV  & 0.9  & p & G & 78      & G & \nodata & \nodata & 2.56 \\
    \hline
\end{tabular}
\end{table*}
\end{center}

\begin{figure}
 \resizebox{0.95\hsize}{!}{\includegraphics[angle=270]{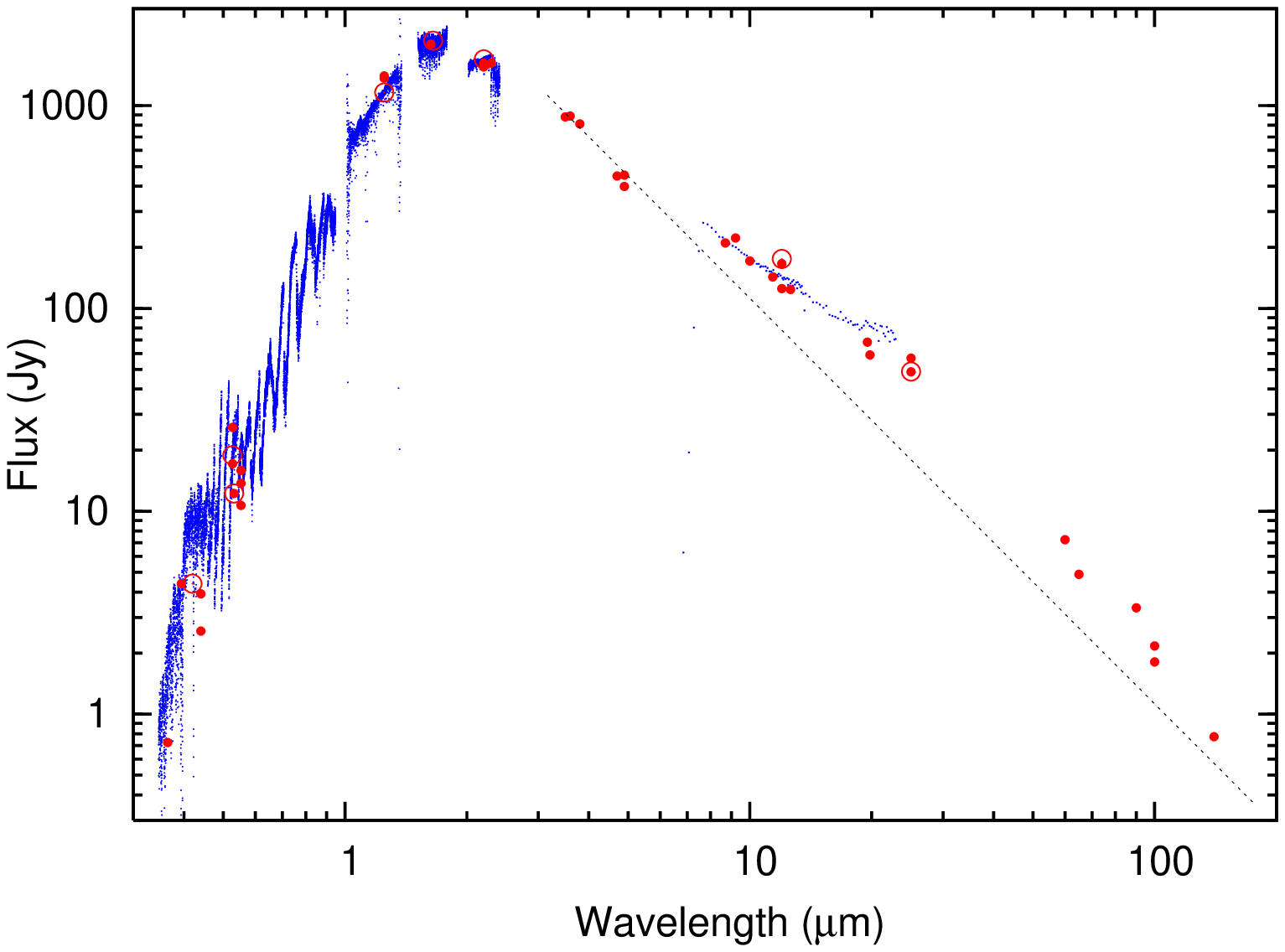}}
 \caption{SED of EU Del. Large red points show literature photometry, small blue points show literature spectra. The black line indicates the Rayleigh--Jeans tail expected for a dustless or `naked' star. References are listed in the text.}
 \label{EUDelFig}
\end{figure}

Figure \ref{GiantXSFig} shows that scatter on the giant branch generally decreases as we approach the AGB tip, due to increased sensitivity on brighter sources. We can also see the substantial number of sources which have scattered to lower temperatures and higher values of $E_{\rm IR}$ (also Figure \ref{ItaXSFig}), indicating reprocessing of optical light be circumstellar dust. The amount of excess around these stars can be correlated with their dust-production rate and hence their mass-loss rate. By identifying and characterising individual stars which lie above the general scatter in Figure \ref{GiantXSFig}, and including optically-obscured sources missed by \emph{Hipparcos}, we can make a theoretically complete census of dust producing stars within 300 pc.

While that is beyond the scope of this paper, we do report on giant stars which are observed to have significant excess. We define this by $E_{IR} > 2.65-\log(L)/3$ (see also Figure \ref{GiantXSFig}), which is chosen to identify excesses of $\gtrsim 2 \sigma$ at all luminosities.  We list these stars in Tables \ref{GiantsTable} (luminous stars, above 850 L$_\odot$) and \ref{Giants2Table} (stars below 850 L$_\odot$). In these tables, we have also listed common names and spectral classification from SIMBAD, and variability information, sourced from either the GCVS (\citealt{SDZ+06}; denoted `G') or \emph{Hipparcos} (denoted `H') catalogues. Variability periods and amplitudes (either in the \emph{Hipparcos} band, denoted `H'; Johnson V-band, denoted `V'; or as a photographic measurement, denoted `p';) are shown where available. Variability types are listed as follows: Mira = Mira variable; SRV = semi-regular variable; Irr = irregular variable; V = unclassified variable; Bin = (eclipsing) binary; None = no appreciable variability. The spectral classifications of any LRS and SWS spectra of these sources are also given (\citealt{SP98,SLMP98,SKPS03}; based on the classification method of \citealt{KSPW02}).

Table \ref{GiantsTable} contains some well-known targets, which are known to have substantial mass loss. The common SE classification of the LWS and SWS spectra shows that most of these stars are known to have silicate features in their spectra and are therefore dust producers. Others, such as SS Lep, $\epsilon$ Car and $\alpha$ Her, are known to be binaries, therefore their SEDs may not be well-repesented by a single blackbody (most giant stars, however, should outshine any companions at all wavelengths: note that $\alpha$ Her in particular is known to be dust producing; \citealt{THWT07}). Several others without LRS or SWS spectra are likely to be mass-losing stars, but without infrared spectra it is difficult to tell.

One outstanding example is present, however: that of EU Del (HIP 101\,810). Figure \ref{EUDelFig} shows its SED. Literature photometry for this SED comes from the \emph{Hipparcos}, \emph{Tycho}, 2MASS, \emph{AKARI} and \emph{IRAS} catalogues already mentioned. Further photometry was sourced from Mermilliod's catalogue of homogeneous means ($UBV$; VizieR online data catalogue II/168), the Carlsberg Meridian Catalogue ($V$; \citealt{CMC99}), the catalogue of infrared observations, including the Revised AFGL catalogue (4--20 $\mu$m; \citealt{PM83,GSPM93}), \emph{DIRBE} (2.2--100 $\mu$m; \citealt{PSK+10}), and the \emph{AKARI} FIS catalogue (65, 90 and 140 $\mu$m; VizieR online data catalogue II/298). Literature spectroscopy for the optical \citep{VGR+04}, $J$-band \citep{WMHE00}, $H$-band \citep{MEHS98} and $K$-band \citep{WH97} are also shown, as is the \emph{IRAS} LRS spectrum from \citet{SP98}. \citet{SP98} classify EU Del as a naked star, as it shows no silicate feature. However, it is found to have substantial excess in both the \emph{IRAS} 12- and 25-$\mu$m bands and the infrared spectrum clearly shows a rise above a blackbody toward longer wavelengths. \citet{WSP+11} place the star at a metallicity of [Fe/H] = --1, making it reminiscent of the featureless excesses we have previously found in metal-poor globular cluster giant stars \citep{MSZ+10,MBvLZ11,MvLS+11}. We have previously attributed this to metallic iron dust on chemical and mineralogical grounds, however it is spectrally indistinguishable from amorphous carbon dust and silicate dust composed primarily of large grains (cf.\ \citealt{HA07}; \citealt{NTI+12}). EU Del may therefore be a unique nearby testbed in which to determine which dust species is causing these unexplained featureless excesses.

Table \ref{Giants2Table} contains a few stars with suspected infrared excess, which we examine more closely, in order to identify the luminosity at which dust formation (as traced by infrared excess) begins. We investigate these individually here.
\begin{list}{\labelitemi}{\leftmargin=1em \itemsep=0pt}
\item AC Cet: Considerable excess exists in the \emph{AKARI} 9- and 18-$\mu$m bands, and the \emph{IRAS} 12- and 25-$\mu$m bands. \citet{CGSW04} note that there is another evolved star within the \emph{IRAS} PSF, but the excess in \emph{AKARI} suggests that AC Cet does indeed have circumstellar dust. \citet{KVS97} classify the source as having a class `C' LRS spectrum, denoting a 11-$\mu$m SiC feature, however this feature is tentative in this source, at best.
\item RU Pup: \emph{AKARI}, \emph{WISE} and \emph{IRAS} data all show considerable excess at wavelengths longer than 4 $\mu$m. There is significant scatter in the optical photometry, which leads to a poor estimation of the temperature and luminosity for this star. This may be partly due to its carbon-richness and partly due to its variability. \citet{BC05} place this star via two means at 2680 or 2875 K and 455 or 610 pc, which makes it considerably cooler and more luminous (1715 or 3649 L$_\odot$) than we model. This is corroborated by its long period (cf.\ \citealt{ITM+04}). It therefore probably suffers from the pulsation-induced distance errors we describe in Section \ref{FitSect}.
\item RS Cam: This star also shows excess in the \emph{AKARI}, \emph{WISE} and \emph{IRAS} data. Its short period suggests its luminosity is correctly determined (cf.\ \citealt{ITM+04}). The LRS spectrum shows weak silicate emission \citep{KVS97}.
\item HR 4091: This source is modelled using \emph{DIRBE} and \emph{IRAS} data only. It shows marginal excess between 4 and 21 $\mu$m, and considerable excess in the 25-$\mu$m \emph{IRAS} band. It is at very low Galactic latitude ($b = -0.5^\circ$) and thus suffers from considerable contamination from surrounding sources. We therefore do not believe this excess is real.
\item TZ Car: The 8- to 25-$\mu$m data for this star show considerable excess. Reprocessing of the \emph{Hipparcos} data by \citet{KPJ01} suggest the distance for this star is roughly correct. At a Galactic latitude of $b = -5.8^\circ$, TZ Car may suffer from some extinction, but it is likely that the excess and parameters are sufficiently correct to say that this star is losing mass.
\item RU Crt: A known mass-losing star, this star shows moderate excess in the \emph{AKARI} bands and substantially more excess in the \emph{IRAS} bands. At 132 pc, it is possible that more-extended emission is missed by \emph{AKARI}: sources up to roughly this distance may have some emission outwith the \emph{AKARI} beam at 25 $\mu$m (see Section \ref{GroenSect}).
\item HD 105\,822: This star is in a region of high projected stellar density ($b = -5.7^\circ$). The amount of excess for this star is inconsistent across the infrared data, varying among the surveys and bands. A dubious $J$-band flux probably suggests more infrared excess than is truly present. The apparent infrared excess in this case is likely due to source blending and confusion, coupled with poor-quality short-wavelength photometry.
\item 68 UMa: This star is mistakenly classified as excessive due to a mismatch between the \emph{Hipparcos/Tycho} magnitudes and those from the SDSS, which differ by approximately a magnitude, despite no variability being detected by \emph{Hipparcos}. This has lead to an excess being determined incorrectly. By using a variety of combinations of photometry, we estimate that a correct temperature and luminosity of around 4000 K and 400 L$_\odot$ would me more appropriate for this source, and that it has no substantial reddening.
\item HR 5464: The determination of infrared excess for this star is based solely on \emph{IRAS} data, in which excess is relatively weak (86\% at 12 $\mu$m, 50\% at 25 $\mu$m). \emph{DIRBE} data suggest there is little or no infrared excess for this source.
\item BD Peg: A known mass-losing star, \citet{KVS97} note that silicate emission is present in this object and \citet{GSPM93} confirm its infrared excess.
\end{list}
With the exception of the carbon star TZ Car, we therefore find no detectable dust production by any object below the luminosity of RU Crt (681 L$_\odot$). We therefore conclude that this represents the luminosity at which dust production by AGB stars begins in earnest in the local neighbourhood. This corroborates very well with the $\approx$700 L$_\odot$ we have previously found in Galactic globular clusters \citep{MvLD+09,BMvL+09,MBvL+11,MvLS+11}.

\subsection{Comparison to Groenewegen (2012)}
\label{GroenSect}

We now turn our attention to the work of \citet{Groenewegen12}. This paper identifies several low-luminosity \emph{Hipparcos} stars with infrared excess which Groenewegen attributes to weak dust emission. These stars have much lower lumionsities (50--350 L$_\odot$) than those we find excess around, as Groenewegen was examining RGB stars for dust excesses much smaller than we have deemed accurately determinable in this work.

The stars from \citet{Groenewegen12} which have survived our data quality cuts are shown on Figure \ref{GiantXSFig}. Groenewegen purposely targetted stars with low ($V-I$) colours, thus his sample does not include stars from with high values of $E_{\rm IR}$. It is notable, however, that all 52 stars common to our datasets lie within the scatter of points with no unusual infrared excess. Also, Groenewegen's dusty stars appear to have no more infrared excess than his dustless stars.

The reasons behind this are not immediately obvious, but can be understood by examining the subtle differences between our analyses. The most striking of these is the choice of input data. We have included data from \emph{WISE}, \emph{AKARI} (9 and 18 $\mu$m) and \emph{IRAS} (12 and 25 $\mu$m). Groenewegen includes data from \emph{AKARI} (including the far-infrared 60- and 90-$\mu$m FIS bands) and \emph{IRAS} (including the far-infrared 60- and 100-$\mu$m bands).

The choice of whether or not to include the far-infrared data is a balance of gaining sensitivity to cold dust and acquiring systematic errors due to contamination in the line of sight. The issue with the \emph{AKARI} FIS and \emph{IRAS} data we have excluded is the beam size (37$^{\prime\prime}$ and 39$^{\prime\prime}$ for \emph{AKARI} FIS 60 and 100 $\mu$m, respectively\footnote{From the \emph{AKARI} FIS data user manual, version 1.3: \url{http://www.ir.isas.jaxa.jp/ASTRO-F/Observation/IDUM/FIS_IDUM_1.3.pdf}}; $4.5^\prime \times 0.7^\prime$ for \emph{IRAS} 12$\mu$m; \citealt{MDL05}). These are much larger than the beam sizes for the other infrared data (5.6$^{\prime\prime}$ for \emph{AKARI} IRC 8.6 $\mu$m\footnote{From the \emph{AKARI} NIR data user manual, version 1.3: \url{http://www.sciops.esa.int/SA/ASTROF/docs/IRC_IDUM_1.3.pdf}}; $7.36^{\prime\prime} \times 6.08^{\prime\prime}$ for \emph{WISE} 11.6$\mu$m \footnote{From the \emph{WISE} preliminary release explanatory supplement: \url{http://wise2.ipac.caltech.edu/docs/release/prelim/expsup/sec4_5c.html}}) and much larger than the optical photometry (typically 1--2$^{\prime\prime}$). A large point-spread function full-width half-maximum (PSF FWHM) means there is a substantial issue with contamination from unrelated sources in the line of sight, from diffuse background emission, or from ISM headed by the star. Equally, if a giant star has a spatially-extended wind, this may be missed by only considering data with a small PSF. Groenewegen has been careful to exclude sources with strong cirrus contamination, which is the main contaminant in the \emph{IRAS} photometry. However, his exclusions are based on the 100-$\mu$m images, whereas ecliptic dust is a greater contaminant at 25 and 60 $\mu$m. These sources would therefore not be identified as contaminated.

On examining the individual stars which \citet{Groenewegen12} claims are dusty, we find that the comparatively-large PSF size of \emph{IRAS} and (in some cases) \emph{AKARI} FIS appears the primary cause of the difference between our datasets. Typically, the \emph{IRAS} 12- and 25-$\mu$m flux is systematically in excess of the modelled stellar photosphere compared to the smaller-PSF \emph{AKARI} IRC and \emph{WISE} photometry. As an example, we model HIP 44\,126 (FZ Cnc) to have moderate excess in \emph{IRAS} (52\% and 35\% at 12 and 25 $\mu$m), but little excess in \emph{WISE} (8\% and 6\% at 11.6 and 22.1 $\mu$m). By only taking the \emph{IRAS} data, Groenewegen naturally models this star as having reasonable infrared excess. In this particular instance, the contaminating source can be clearly identified as poorly-subtracted emission from warm dust in the ecliptic plane in the original \emph{IRAS} photometry. The same is true of HIP 53\,449, though here is the \emph{AKARI} FIS data that suffers from contamination from the ecliptic.

Not all of Groenewegen's dusty sources can be explained so easily, however. HIP 67\,605 and 67\,665 (AW CVn) are both identified as dusty by \citet{Groenewegen12}. They lie quite close to each other (15$^\prime$ apart) but are resolved in the \emph{IRAS} images by several beam widths. They do not suffer from substantial contamination. They are covered by \emph{AKARI} and \emph{IRAS}, but not the \emph{WISE} preliminary catalogue. Both sources have excess at \emph{IRAS} 12 and 25 $\mu$m but not \emph{AKARI} 9 and 18 $\mu$m. Like the majority of Groenewegen's targets, these stars lie at around 200 pc. At this distance the \emph{AKARI} 8.6-$\mu$m PSF has a FWHM of 1120 AU, or 2000--4000 stellar radii. Assuming the dust temperature approximately follows a $T^4 \propto R^2$ law, and a stellar temperature of $\approx$3700 K, this implies that dust falling within the \emph{IRAS} beam but outwith the \emph{AKARI} IRC beam should emit with a peak wavelength of $\lambda \gg 35$ $\mu$m. It should therefore not emit significantly at 8--18 $\mu$m to cause the discrepency between the \emph{AKARI} NIR and \emph{IRAS} 12-$\mu$m fluxes.

Nevertheless, the infrared excess Groenewegen finds may still be real, and still be related to a wind eminating from the star. Two situations may cause this. The first case is that a cooler, detached shell surrounds the star and emits only at longer (60--100 $\mu$m) wavelengths (cf.\ Y CVn; \citealt{LGlB07}). In the second case, the emission would be produced not by the star, but instead interstellar dust swept up in a bow shock around the astropause \citep{WZOS07}. This has been seen in other nearby AGB stars \citep{USS+06,USY+10,LBG+10} and could be the source of the excess emission at longer (60--100 $\mu$m) wavelengths that Groenewegen finds in several cases. 

We also acknowledge that variability may also play a r\^{o}le in this analysis. Neither Groenewegen's nor our determinations of infrared excess take into account the variability of stars. As \citet{Groenewegen12} uses some different optical data to us, we may find that some stars in both studies appear to have infrared excess simply because their optical photometry was observed when the star was at photometric minimum. Conversely, excess may be missed if observations were carried out at photometric maximum.

\citet{Groenewegen12} and this work probe subtly different datasets with subtly different techniques. It should therefore not be surprising that we find different results, though we would argue that our analysis should be better suited to finding ongoing dust production by stars. On the basis of the above discussion, we advise caution when investigating small infrared excesses in such cases and note the benefits of phase-matched, high-resolution infrared photometry (see also \citealt{SMM+10,MBvL+11,MSS+12}).


\section{Conclusions}
\label{ConcSect}

In this work, we have demonstrated the use of spectral energy distribution fitting to determine the fundamental parameters of the \emph{Hipparcos} star sample. We have further used this information to quantify excess flux over the entire optical, and near- and mid-infrared region of each SED. We have combined these excesses to determine those stars showing an excess of infrared flux, and cross-correlated literature identifications to examine the cause of that excess over different regions of the H--R diagram, comparing our results to the key studies of \citet{IMI+10} and \citet{Groenewegen12}. We find we cannot reproduce the infrared excess and dust production claimed by the latter paper.

Our analysis has focussed on the \emph{Hipparcos} data catalogue: data which is now over 20 years old and, despite showing its age, provides the best estimate of distances to nearby stars we have. The launch of \emph{Gaia}, and the completion of further all-sky surveys such as Pan-STARRS, SDSS and \emph{WISE}, will allow a similar analysis to be performed on many times more objects. Automated techniques, building on the kind demonstrated here, will be necessary to analyse and classify the objects which come from these surveys, in order to gain a full and comprehensive understanding of our corner of the Galaxy and its inhabitants.


\section*{Acknowledgements}

We thank Martin Groenewegen for his invaluable input into the comparison with his published works and his help in resolving the differences between our results.

This research has made extensive use of the SIMBAD database, VizieR catalogue access tool and Aladin, operated at CDS, Strasbourg, France.

This research has made use of the NASA/IPAC Infrared Science Archive, which is operated by the Jet Propulsion Laboratory, California Institute of Technology, under contract with the National Aeronautics and Space Administration.

The \emph{Hipparcos}/Tycho catalogues are a result of the \emph{Hipparcos} space astrometry mission, undertaken by the European Space Agency.

Funding for SDSS-III has been provided by the Alfred P.\ Sloan Foundation, the Participating Institutions, the National Science Foundation, and the U.S.\ Department of Energy. The SDSS-III web site is http://www.sdss3.org/.

The DENIS project has been partly funded by the SCIENCE and the HCM plans of the European Commission under grants CT920791 and CT940627. It is supported by INSU, MEN and CNRS in France, by the State of Baden-W\"urttemberg in Germany, by DGICYT in Spain, by CNR in Italy, by FFwFBWF in Austria, by FAPESP in Brazil, by OTKA grants F-4239 and F-013990 in Hungary, and by the ESO C\&EE grant A-04-046. Jean Claude Renault from IAP was the Project manager.  Observations were carried out thanks to the contribution of numerous students and young scientists from all involved institutes, under the supervision of  P. Fouqu\'e, survey astronomer resident in Chile.

This publication makes use of data products from the Two Micron All Sky Survey, which is a joint project of the University of Massachusetts and the Infrared Processing and Analysis Center/California Institute of Technology, funded by the National Aeronautics and Space Administration and the National Science Foundation.

This research made use of data products from the Midcourse Space Experiment.  Processing of the data was funded by the Ballistic Missile Defense Organization with additional support from NASA Office of Space Science.

This publication makes use of data products from the \emph{Wide-field Infrared Survey Explorer}, which is a joint project of the University of California, Los Angeles, and the Jet Propulsion Laboratory/California Institute of Technology, funded by the National Aeronautics and Space Administration.

This research is based on observations with \emph{AKARI}, a JAXA project with the participation of ESA.


\appendix
\section{Removal of bad data}

We describe in this Appendix the sequentially-applied cuts we use to remove bad data from our catalogue.

\subsection{Cut \#1}
\label{Cut1Sect}

The first cut was designed to remove extremely cool sources and sources with distinctly double-peaked SEDs from the sample. These tend to be heavily-enshrouded or heavily-extincted objects which we cannot accurately model.

Stars were removed from the main catalogue if they had at least two mid-infrared (3.5--25 $\mu$m) bands brighter than all their optical/near-IR ($u^\prime$--$K_{\rm s}$-band) data. This cut removed 241 objects from the catalogue, of which 98 objects have no $IJHK_{\rm s}$-band data, leaving 109\,436.

\subsection{Cut \#2}
\label{Cut2Sect}

The second cut was designed to remove unphysically low values from the \emph{WISE} data, which are much too faint to come from a stellar object detectable by \emph{Hipparcos}. A \emph{WISE} photometric datum was removed from the combined catalogue if its flux was below 100 $\mu$Jy. In this way, \emph{WISE} photometry was deleted from 281 objects.

\subsection{Cut \#3}
\label{Cut3Sect}

The third cut acts to remove more underluminous $W_1$ and $W_2$ data. To assist in this, we define two fluxes, $F_{10}$ and $F_{20}$: $F_{10}$ is the 10-$\mu$m flux defined by (in order of preference) the $W_3$, \emph{AKARI} [9] or \emph{IRAS} [12] flux, and $F_{20}$ is defined similarly by the $W_4$, \emph{AKARI} [18] or \emph{IRAS} [25] flux.

This cut applies to all stars where $F_{10} > F_{20}$ and stars where no $F_{10}$ measurement exists. This requirement prevents the selection of dusty sources where the SED reaches a minimum between 3 and 8 $\mu$m but where dust emission is insufficient to be picked up by Cut \#1. $W_1$ and/or $W_2$ are removed if it their fluxes are less than 40\% of both the 2MASS (or DENIS, where substituted) $K_{\rm s}$-band and the $F_{10}$ fluxes. Where no measure of $F_{10}$ is available, $F_{20}$ is used instead.

The cut removes \emph{WISE} data from 273 objects.

\subsection{Cut \#4}
\label{Cut4Sect}

The fourth cut removes more underluminous $W_1$ data. This cut detects objects where the flux in $W_1$ is $<$40\% of the flux in $W_2$, provided the flux in $W_2$ is: (1) non-zero, (2) greater than the flux in $W_3$ and (3) less than the flux in $K_{\rm s}$ (where $W_3$ and $K_{\rm s}$ fluxes exist). The Cepheid X Sgr (HIP\,87072) is excluded from this cut as a special case. This removes 38 bad $W_1$ datapoints from the catalogue.

At this point, sources were also removed from the catalogue if they had data in five or fewer of the observed bands. This removed 1479 objects, leaving 107\,957. Most of the deletions were either detected by \emph{Hipparcos} and \emph{IRAS}, or \emph{Hipparcos} and SDSS.

\subsection{Cut \#5}
\label{Cut5Sect}

The fifth cut is designed to remove single bad photometric data points. It operates on all filters shortward of 8.6 $\mu$m, except the \emph{Hipparcos}/\emph{Tycho} data.

For each \emph{Hipparcos} object, we determine the worst-fitting filter out of those listed in the previous paragraph, i.e.\ the filter with the greatest value of $R$ or $1/R$. If this value is five times greater than the next largest $R$ or $1/R$, it is removed. For example, if $J$ and $K_{\rm s}$ are the two worst-fitting filters, $R_J = 51$ and $R_Ks = 10$ then the $J$-band datum will be removed, whereas if $R_J = 49$ it will not.

This cut removes 78 bad datapoints from the catalogue. Of these, 60 are DENIS $I$-band fluxes.

\subsection{Cut \#6}
\label{Cut6Sect}

The sixth cut was done manually to remove five stars where IR photometry is clearly confused due to blending, variability or background. These stars are HIP 60782, 80057 and 88267 (all binaries), and HIP 82850 and 82611. This left 107\,952 unique catalogued objects.

\subsection{Cut \#7}
\label{Cut7Sect}

The seventh cut removes \emph{MSX} $B_1$ and/or $B_2$ data when these data have a higher flux than than the $K_{\rm s}$ filters and one of the $W_3$, $AKARI$ [9] or $IRAS$ [12] filters. While in principle this could remove points from SEDs peaking between 2.2 and 12 $\mu$m, no objects seem to be affected by this. Dustless stars tend to peak at wavelengths shorter than 2.2 $\mu$m, while dusty stars with double-peaked SEDs tend to have their second peak at wavelengths longer than 12 $\mu$m. Data was removed from 244 objects, though this affected only fraction of these, as $MSX$ data is only used when WISE data is unavailable. There were 13 objects removed during this stage because they had insufficient photometry, leaving 107\,939 unique objects.

\subsection{Cut \#8}
\label{Cut8Sect}

The eighth cut is designed to remove underluminous SDSS data from saturated sources. SDSS $g$- and/or $r$-band data are removed if they are recorded to have less flux than the $Hipparcos$ and $Tycho$ $B_T$ and $V_T$ fluxes. This affects 7058 objects, though most of these were already rejected in \S\ref{DataSect}.

An additional 224 points were manually removed from 147 objects, mainly consisting of errant $W_1$ and $W_2$ fluxes. A further 39 objects were rejected for having insufficient data, leaving 107\,900 unique objects.

\subsection{Cut \#9}
\label{Cut9Sect}

The ninth cut repeats cut \#5, removing datapoints which have a goodness-of-fit more than a factor 2.5 (instead of 5) worse than the next worst-fitting point. This cut was performed in two interations, the first removing a point from 647 objects, the second removing a point from 59 objects. A final 120 points from 112 objects were edited out by hand. Following this, objects with less than five photometric points at $<$8 $\mu$m were rejected, leaving 107\,619 unique objects in the final catalogue.



\label{lastpage}

\end{document}